\documentclass[pra,twocolumn,superscriptaddress,preprintnumbers,amsmath,amssymb]{revtex4}
\usepackage{}

\usepackage{amsfonts}
\usepackage{amsmath}
\usepackage{amssymb}
\usepackage{amstext}
\usepackage{amsthm}
\usepackage{bm}
\usepackage{bbm}
\usepackage{booktabs}
\usepackage{braket}
\usepackage{color}%
\usepackage{dcolumn}
\usepackage{dsfont}
\usepackage{epsfig,graphicx}
\usepackage{float}
\usepackage{graphicx}
\usepackage{mathrsfs}
\usepackage{mathtools}
\usepackage{mleftright}
\usepackage{times}
\usepackage{txfonts}
\usepackage[utf8]{inputenc}

\usepackage{hyperref}
\hypersetup{colorlinks=true,linkcolor=blue,citecolor=blue,urlcolor=blue}

\setcitestyle{open={[},close={]},citesep={,\!},numbers}

\newcommand{\iden}{\mathds{1}}
\newcommand{\tr}{\mathrm{tr}}

\begin{document}

\title{Off-resonant preservation and generation of imaginarity in distributed scenarios}

\author{Si-Min Wang}
\address{School of Science, Xi'an University of Posts and Telecommunications, Xi'an 710121, China}

\author{Ming-Liang Hu}
\email{mingliang0301@163.com}
\address{School of Science, Xi'an University of Posts and Telecommunications, Xi'an 710121, China}

\author{Heng Fan}
\email{hfan@iphy.ac.cn}
\affiliation{Institute of Physics, Chinese Academy of Sciences, Beijing 100190, China}
\affiliation{School of Physical Sciences, University of Chinese Academy of Sciences, Beijing 100190, China}
\affiliation{Beijing Academy of Quantum Information Sciences, Beijing 100193, China}

\begin{abstract}
We study the nonlocal advantage of quantum imaginarity (NAQI) and distillable imaginarity of assistance (DIA), which treat imaginarity as a resource in distributed scenarios. For two qubits interacting with a lossy cavity, it is shown that both the NAQI and DIA can be well preserved for long times in the presence of large and symmetric detuning between the qubits and the cavity. Moreover, the off-resonant interaction generates a high degree of NAQI and DIA from the initial product states of two qubits having the same detunings and unequal couplings to the cavity. Based on the effective coupling of the qubits induced by the cavity mode, we explain the physical mechanism underlying the validity of this strategy. Our findings shed light on the role that off-resonant interactions have in the efficient control of imaginarity in distributed scenarios.
\end{abstract}

\pacs{03.65.Ta, 03.65.Ud, 03.67.-a
\quad Keywords: resource theory, imaginarity, non-Markovian dynamics}

\maketitle

\section{Introduction} \label{sec:1}
Imaginary numbers are essential to the description of physical reality, and the necessity and usefulness of imaginary numbers in accurately formulating quantum theory have also been demonstrated experimentally and theoretically \cite{real1,real2,real3,real4,real5}. In view of this, it is instructive to quantify the imaginarity in a  state. In a similar vein to the resource theories of entanglement \cite{QE0,QE1}, coherence \cite{coher,Plenio,Hu}, superposition \cite{Rsuper}, and steering \cite{Rsteer}, Hickey and Gour introduced a resource theory of imaginarity in which a state with imaginary terms in a given basis is regarded as resourceful \cite{ima_jpa}. On the basis of this framework, a number of imaginarity measures have been proposed. Some well-defined ones include the trace norm of imaginarity \cite{ima_jpa}, the robustness of imaginarity \cite{ima_jpa,ima_prl,ima_pra}, the geometric imaginarity \cite{ima_pra,ima_njp}, the relative entropy of imaginarity and the weight of imaginarity \cite{ima_qip}, the imaginarity measures by the convex roof construction and the least imaginarity of the input pure states under real operations \cite{ima_dus}, the imaginarity measure by the stabilized quantum optimal transport cost \cite{ima_lin}, the $\epsilon$ measure of imaginarity \cite{ima_epsilon}, the generalized quantum Jensen-Shannon divergence of imaginarity \cite{ima_suny}, as well as some entropic measures of imaginarity \cite{ima_pla,ima_aqt,ima_maxmin,ima_ctp}.

The past few years have also witnessed a surge of interest in studying quantum imaginarity from other perspectives; in particular, the resource theory provides a useful framework to identify its interrelation with coherence \cite{ima_qc1,ima_qc2,ima_qc3}, entanglement \cite{ima_ent1,ima_ent2}, and quantum speed limit \cite{ima_speed}. The operational interpretation of imaginarity \cite{ima_prl,ima_dus,ima_maxmin,ima_erase}, its witness \cite{ima_wit1,ima_wit2}, as well as the imaginarity of Gaussian states \cite{ima_gauss1,ima_gauss2} and quantum channels \cite{ima_channel} have also been discussed. The recognition that imaginarity is a physical resource makes it more promising in quantum information science, for example, many efforts have been dedicated to understanding its role in quantum tasks such as state conversion and state discrimination \cite{ima_prl,ima_pra}, hiding and masking \cite{ima_mask1,ima_mask2}, multiparameter estimation \cite{ima_estim}, and broadcasting \cite{ima_broad1,ima_broad2}.

While quantum imaginarity is usually defined with respect to a single-partite system \cite{ima_jpa}, one can further consider its distribution property among different subsystems of a bipartite or multipartite system, perhaps with the help of local operations and classical communication (LOCC). In this way the nonlocal advantage of quantum imaginarity (NAQI) for a two-qubit state $\rho_{AB}$ was introduced \cite{naqi}. Following the similar steps as defining the nonlocal advantage of quantum coherence \cite{naqc1,naqc2,naqc3}, the NAQI is defined by considering degree of the steered imaginarity on qubit $B$ under mutually unbiased bases (MUBs); specifically, if the steered imaginarity achieved by LOCC violates the imaginarity complementarity relation (see Sec. \ref{sec:2} for details) for a single qubit, it is said that there is NAQI in $\rho_{AB}$. The NAQI is stronger than quantum steering in the sense that any $\rho_{AB}$ with NAQI is steerable but not vice versa \cite{naqi}. Apart from NAQI, another paradigm for studying imaginarity in distributed scenarios is by local quantum-real operations and classical communication (LQRCC), which is termed assisted imaginarity distillation \cite{ima_cp}, in analogy to assisted coherence distillation \cite{dist_coher}. The objective here is to localize an optimal degree of imaginarity on $B$ for $\rho_{AB}$ shared in prior, and the corresponding distillable imaginarity of assistance (DIA) describes the degree of  imaginarity the party $B$ can gain when getting assistance from a remote collaborative party $A$.

The resource theoretic framework of imaginarity also opens perspective for a quantitative analysis of quantum imaginarity in open systems, and along this line, the resilience of single-qubit imaginarity to the paradigmatic bit flip, phase flip, and amplitude damping noises have been discussed \cite{ima_aqt,ima_dy1}, and the freezing of imaginarity under real operations has also been investigated \cite{ima_free}. Despite these advances, the control of imaginarity in multipartite systems under realistic scenarios remains largely unexplored. To fill this gap we consider in this paper the NAQI and DIA for two qubits interacting with a common reservoir formed by the electromagnetic field inside a lossy cavity. This consideration treats imaginarity as a physical resource in distributed scenarios, which may correspond to, e.g., the case of a remote party on which imaginarity is needed as a resource. We focus on two problems: (\romannumeral+1) the preservation of NAQI and DIA for the initial Bell states, and (\romannumeral+2) the generation of NAQI and DIA from the initial product states. We will show that this objective could be achieved by tuning detunings of the qubits from the fundamental frequency of the cavity and the coupling strengths of the qubits to the cavity. We will also explain the origin of the large off-resonant control strategy by considering effective coupling of the qubits mediated by the cavity mode through exchanging virtual photons.

The rest of this paper is organized as follows. In Sec. \ref{sec:2} we recall briefly the preliminaries of NAQI and DIA, and in Sec. \ref{sec:3} we present solution of the physical model. Secs. \ref{sec:4} and \ref{sec:5} are devoted to analyzing preservation and generation of NAQI and DIA, respectively. Finally, we conclude this paper with a short summary and discussion in Sec. \ref{sec:6}.

\section{NAQI and DIA} \label{sec:2}
In this section we recall the preliminaries related to NAQI and DIA. We first recall the framework for quantifying imaginarity in a quantum state described by the density operator $\rho$. Similar to the established resource theories of entanglement \cite{QE0,QE1} and coherence \cite{Plenio,Hu}, the resource theory of imaginarity also comprises three ingredients: the free states, the free operations, and some axiomaticlike prerequisites that an imaginarity measure $\mathcal{I}(\rho)$ should satisfy \cite{ima_jpa,ima_prl,ima_pra}. The free states are real states, which are described by those $\rho$ with $\langle i|\rho|j\rangle \in \mathbb{R}$ ($\forall i, j$) in the reference basis $\{|i\rangle\}$, the free operations are real operations, that is, quantum operations $\Lambda[\rho]=\sum_l K_l\rho K_l^\dag$ with the Kraus operators $\{K_l\}$ satisfying $\langle i|K_l|j\rangle \in \mathbb{R}$ ($\forall i, j, l$), and the prerequisites for $\mathcal{I}(\rho)$ are as follows: (I1) $\mathcal{I}(\rho)\geqslant 0$, and the equality holds if and only if $\rho\in \mathscr{R}$, with $\mathscr{R}$ being the set of real states; (I2) $\mathcal{I}(\Lambda[\rho]) \leqslant \mathcal{I}(\rho)$ for any real operation $\Lambda$; (I3) $\sum_l p_l \mathcal{I}(\rho_l) \leqslant \mathcal{I}(\rho)$, where $\rho_l= K_l\rho K_l^\dag / p_l$ and $p_l=\tr (K_l\rho K_l^\dag)$; (I4) $\mathcal{I}(\sum_i p_i \rho_i) \leqslant \sum_i p_i \mathcal{I}(\rho_i)$ for any ensemble $\{p_i,\rho_i\}$, where $\{p_i\}$ is the probability distribution.

Within the resource theoretic framework, there is a wealth of imaginarity measures being proposed in the past few years \cite{ima_jpa,ima_prl,ima_pra,ima_njp,ima_qip,ima_dus,ima_lin,ima_epsilon,ima_suny,ima_pla,ima_aqt, ima_maxmin,ima_ctp}. In this work we consider the well-defined trace norm of imaginarity $\mathcal{I}_{tr}(\rho)$ \cite{ima_jpa}, the geometric imaginarity $\mathcal{I}_g(\rho)$ \cite{ima_pra,ima_njp}, and the relative entropy of imaginarity $\mathcal{I}_{re}(\rho)$ \cite{ima_qip}, which are defined, respectively, by
\begin{equation} \label{eq2-1}
\begin{aligned}
 & \mathcal{I}_{tr}(\rho)= \min_{\sigma\in\mathscr{R}} \|\rho-\sigma \|_1, ~
   \mathcal{I}_{re}(\rho)= \min_{\sigma\in\mathscr{R}} S(\rho\|\sigma), \\
 & \mathcal{I}_g(\rho)= \min_{\{p_l,|\psi_l\rangle\}}\sum_l p_l \mathcal{I}_g(|\psi_l\rangle),
\end{aligned}
\end{equation}
where $\|X\|_1 = \tr(X^\dag X)^{1/2}$ represents the trace norm, $S(\rho\|\sigma)= \tr(\rho\log_2\rho)- \tr(\rho\log_2\sigma)$ is the relative entropy, $\{p_l,|\psi_l\rangle\}$ is the pure state decompositions of $\rho=\sum_l p_l |\psi_l\rangle\langle\psi_l|$, and $\mathcal{I}_g(|\psi_l\rangle)= 1-\max_{|\varphi\rangle\in\mathscr{R}} |\langle\varphi|\psi_l\rangle|^2$. These imaginarity measures are favored for their compact analytical solutions, which are given by
\begin{equation} \label{eq2-2}
\begin{aligned}
 & \mathcal{I}_{tr}(\rho)= \frac{1}{2}\|\rho-\rho^T \|_1, ~
   \mathcal{I}_{re}(\rho)= S(\rho_R)-S(\rho), \\
 & \mathcal{I}_g(\rho)= \frac{1}{2} \left(1-\sqrt{F(\rho,\rho^T)} \right),
\end{aligned}
\end{equation}
where $\rho^T$ is the transpose of $\rho$, $\rho_R= (\rho+\rho^T)/2$, and $F(\rho,\rho^T)= \big(\tr\sqrt{\rho^{1/2} \rho^T \rho^{1/2}}\big)^2$ is the Uhlmann fidelity. Furthermore, for the robustness of imaginarity $\mathcal{I}_R(\rho)$, one has $\mathcal{I}_R(\rho)= \mathcal{I}_{tr}(\rho)$ \cite{ima_jpa,ima_prl,ima_pra}; thus our results also apply to it.

As the imaginarity measure is intrinsically basis dependent, the sum of the one-qubit imaginarity with respect to the three MUBs $\{\mathcal{M}_i\}$ (the unitary equivalent classes of the eigenbases of the Pauli matrices $\sigma_{1,2,3}$) has an upper bound, that is,
\begin{equation} \label{eq2-3}
 \sum_i \mathcal{I}^{\mathcal{M}_i}_\alpha(\rho) \leqslant \mathcal{I}_{\alpha, \mathcal{B}},
\end{equation}
where $\alpha \in \{tr,re,g\}$, and $\mathcal{I}^{\mathcal{M}_i}_\alpha(\rho)$ is the imaginarity measure with $\rho$ being written in the reference basis $\mathcal{M}_i$. In Ref. \cite{naqi} it has been shown that $\mathcal{I}_{tr,\mathcal{B}}= \sqrt{5}$ and $\mathcal{I}_{re,\mathcal{B}}\approx 2.02685$. For the geometric imaginarity, the upper bound can be obtained as $\mathcal{I}_{g,\mathcal{B}}=1$ (see Appendix \ref{sec:A}).

\begin{figure}
\centering
\resizebox{0.47 \textwidth}{!}{%
\includegraphics{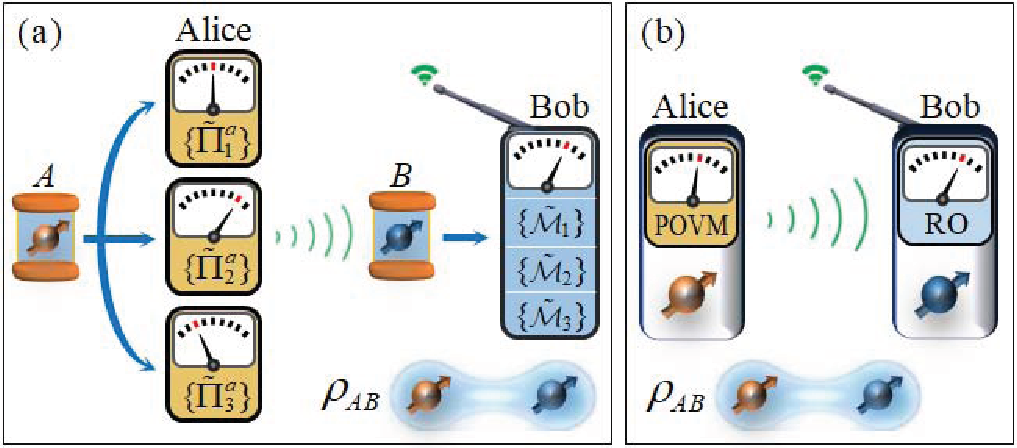}}
\caption{Illustration of the imaginarity steering procedure in distributed scenarios, where the duo Alice ($A$) and Bob ($B$) share a two-qubit state $\rho_{AB}$ in prior. (a) The NAQI on qubit $B$ achieved by LOCC. (b) The assisted imaginarity distillation on qubit $B$ by LQRCC.} \label{fig:1}
\end{figure}

Starting from the above preliminaries, one can introduce the NAQI, which is similar in spirit to the nonlocal advantage of quantum coherence \cite{naqc1,naqc2,naqc3}. As is illustrated in Fig. \ref{fig:1}(a), a two-qubit state $\rho_{AB}$ is shared between Alice ($A$) and Bob ($B$), Alice performs her projective measurements chosen randomly from $\{\Pi_i\}_{i=1}^3$ on qubit $A$ and informs Bob of her measurement choice $\Pi_i$ and outcome $a\in\{+,-\}$. Then the conditional state of Bob will be given by
\begin{equation} \label{eq2-4}
 \rho_{B|\Pi_i^a}= \tr_A [(\Pi_i^a\otimes \iden)\rho_{AB}(\Pi_i^a\otimes \iden)]/p_{B|\Pi_i^a},
\end{equation}
where $p_{B|\Pi_i^a}=\tr[\Pi_i^a \otimes \iden)\rho_{AB}]$ is the probability of the measurement outcome $a$, and $\iden$ is the one-qubit identity operator. The elements of $\Pi_i= \{\Pi_i^\pm\}$ are given by $\Pi_i^\pm = (\iden \pm \bm{n}_i \cdot \bm{\sigma})/2$, where $\bm{\sigma}= (\sigma_1,\sigma_2,\sigma_3)$ and $\bm{n}_i=(\sin\theta_i\cos\phi_i, \sin\theta_i\sin\phi_i, \cos\theta_i)$, with $\theta_i \in [0,\pi]$ and $\phi_i \in [0,2\pi]$ ($\forall i$).

After repeating this process many times, one has the ensemble $\{p_{B|\Pi_i^a},\rho_{B|\Pi_i^a}\}$, the average quantum imaginarity of which is $\sum_a p_{B|\Pi_i^a} \mathcal{I}_\alpha^{\mathcal{M}_i}(\rho_{B|\Pi_i^a})$. By adding up them over the three $\mathcal{M}_i$ and then maximizing over the full sets of $\{\Pi_i\}$ and $\{\mathcal{M}_i\}$, one obtains the sum of the average imaginarities on $B$ with respect to the MUBs, which is given by
\begin{equation} \label{eq2-5}
 \mathcal{N}_\alpha(\rho_{AB})= \max_{\{\Pi_i,\mathcal{M}_i\}} \sum_{i,a} p_{B|\Pi_i^a} \mathcal{I}_\alpha^{\mathcal{M}_i}(\rho_{B|\Pi_i^a}),
\end{equation}
and the elements of $\mathcal{M}_i= \{|\mathcal{M}_i^\pm\rangle\}$ can be written as
\begin{equation} \label{eq2-6}
\begin{aligned}
 & |\mathcal{M}_1^{+}\rangle= \cos(\theta_4/2)|1\rangle+ e^{i\phi_4}\sin(\theta_4/2)|0\rangle, \\
 & |\mathcal{M}_1^{-}\rangle= \sin(\theta_4/2)|1\rangle- e^{i\phi_4}\cos(\theta_4/2)|0\rangle, \\
 & |\mathcal{M}_2^\pm\rangle= \frac{|\mathcal{M}_1^{+}\rangle \pm  |\mathcal{M}_1^{-}\rangle} {\sqrt{2}}, ~
   |\mathcal{M}_3^\pm\rangle= \frac{|\mathcal{M}_1^{+}\rangle \pm i|\mathcal{M}_1^{-}\rangle} {\sqrt{2}},
\end{aligned}
\end{equation}
where $\theta_4\in[0,\pi]$ and $\phi_4\in [0,2\pi]$. For $\rho_{AB}=\rho_{A}\otimes \rho_B$, it is obvious that $\mathcal{N}_\alpha(\rho_{AB}) \leqslant \mathcal{I}_{\alpha,\mathcal{B}}$ as $\rho_{B|\Pi_i^a} \equiv \rho_B$ ($\forall i,a$). But for an entangled $\rho_{AB}$, the case may be different. Thus, a criterion for capturing the NAQI in $\rho_{AB}$ can be established as \cite{naqi}
\begin{equation} \label{eq2-7}
 \mathcal{N}_\alpha(\rho_{AB}) > \mathcal{I}_{\alpha,\mathcal{B}},
\end{equation}
and for the Bell states, $\mathcal{N}_{tr,re}(\rho_{\mathrm{Bell}})= 3$ and $\mathcal{N}_g(\rho_{\mathrm{Bell}})=1.5$. This corresponds to the maximum violation of Eq. \eqref{eq2-3}.

The assisted imaginarity distillation task is sketched in Fig. \ref{fig:1}(b). Here, Alice and Bob also share $\rho_{AB}$, but their goal is to distill as much imaginarity as possible on $B$ via LQRCC, that is, by Alice's positive operator-valued measurements (POVMs) $\{E_i\}$ on $A$ and Bob's real operations $\{\Lambda_i\}$ on $B$, together with the one-way classical communication from Alice to Bob. For the single-shot scenario, the DIA could be assessed by the assisted fidelity of imaginarity given below:
\begin{equation} \label{eq2-8}
 F_a(\rho_{AB})= \max_{\{E_i,\Lambda_i\}} \sum_i p_{B|E_i} F\big(\Lambda_i[\rho_{B|E_i}], |\tilde{+}\rangle\langle\tilde{+}|\big),
\end{equation}
where $p_{B|E_i}=\tr[(E_i\otimes\iden)\rho_{AB}]$, $\rho_{B|E_i}= \tr_A [(E_i\otimes \iden)\rho_{AB}]/p_{B|E_i}$, and $|\tilde{+}\rangle=(|0\rangle+i|1\rangle)/\sqrt{2}$ is the maximally imaginary state.

By rewriting the two-qubit state as $\rho_{AB}= \tfrac{1}{4} \sum_{k,l=0}^3 v_{kl}\sigma_k\otimes\sigma_l$, with $\sigma_0= \iden$ and $v_{kl}= \tr(\rho_{AB}\sigma_k\otimes\sigma_l)$, one can obtain \cite{ima_cp}
\begin{equation} \label{eq2-9}
 F_a(\rho_{AB})= \frac{1}{2}\left(1+\max\{|v_{02}|,|\bm{s}|\}\right),
\end{equation}
where $\bm{s}= (v_{12}, v_{22}, v_{32})$. This indicates that the assisted imaginarity distillation task has a quantum advantage compared to that without Alice's assistance when $|\bm{s}|> |v_{02}|$.

\section{Solution of the model} \label{sec:3}
In this paper we consider a system consisting of two qubits $A$ and $B$ coupled to a common bosonic reservoir in the vacuum initially. The Hamiltonian (in units of $\hbar$) of the system plus the reservoir, in the rotating-wave approximation (RWA), is given by $H = H_{SR}+H_\mathrm{int}$, with
\begin{equation}\label{eq3-1}
\begin{aligned}
 & H_{SR}= \sum_{n=A,B} \omega_n \sigma^{n}_{+} \sigma^{n}_{-} + \sum_k \omega_k b_k^\dagger b_k, \\
 & H_\mathrm{int}= (\alpha_A \sigma^{A}_{+}+\alpha_B \sigma^{B}_{+}) \sum_k g_k b_k + \mathrm{H.c.},
\end{aligned}
\end{equation}
where $\omega_n$ and $\sigma^{n}_{\pm}$ are the transition frequency and inversion operators of qubit $n$ ($n=A,B$), $\omega_k$ and $b_k$ ($b_k^\dagger$) are the frequency and annihilation (creation) operator of the $k$th mode of the reservoir, and its coupling strength to qubit $n$ is $\alpha_n g_k$, with $\alpha_n$ being a dimensionless constant determined by the reservoir at the position of qubit $n$, and following Refs. \cite{common1,common2}, we define $\alpha_T= (\alpha_A^2+\alpha_B^2)^{1/2}$ and $r_n=\alpha_n/\alpha_T$, where $r_n \geqslant 0$ ($\forall n$).

To solve evolution equation of the system, we need to know the spectrum of the reservoir. In this study we consider the reservoir formed by the electromagnetic field inside a lossy cavity. It has a Lorentzian spectral density \cite{common2}
\begin{equation}\label{eq3-2}
 J(\omega)= \frac{W^2}{\pi} \frac{\lambda}{(\omega-\omega_c)^2+\lambda^2},
\end{equation}
where the nonperfect reflectivity of the cavity mirrors induces the Lorentzian broadening of the spectrum, $W$ describes the system-cavity coupling, $\lambda$ is the frequency width of the spectrum and characterizes the cavity loss rate ($1/\lambda$ is the reservoir correlation time and the ideal cavity corresponds to $\lambda\rightarrow 0$), and $\omega_c$ is the central frequency of the cavity field. Throughout this paper we denote by $\delta_n = \omega_n- \omega_c$ ($n=A,B$) the frequency detuning of qubit $n$ from $\omega_c$.

For the $\omega_A = \omega_B$ case, if one considers the initial two-qubit state of the form
\begin{equation}\label{eq3-3}
 |\psi(0)\rangle = c_{10}|10\rangle + c_{20}|01\rangle,
\end{equation}
the ``qubits\,+\,reservoir" state at time $t$ can be written as
\begin{equation}\label{eq3-4}
 |\Psi(t)\rangle = c_1(t) |10\bar{0}\rangle + c_2(t) |01\bar{0}\rangle
                   +\sum_k c_k(t)|00\bar{1}_k\rangle,
\end{equation}
where $|\bar{0}\rangle=\otimes_k |\bar{0}_k\rangle$, and $|\bar{0}_k\rangle$ and $|\bar{1}_k\rangle$ denote, respectively, the states of the reservoir with zero and one excitation in the mode $k$.  After obtaining $c_1(t)$ and $c_2(t)$, $\rho_{AB}(t)$ can be obtained via partial tracing over the reservoir degrees of freedom. Solving the Schr\"{o}dinger equation $i\partial |\Psi(t)\rangle / \partial t =H_{I}(t) |\Psi(t)\rangle$ ($H_I$ is the  Hamiltonian in the interaction picture), one has \cite{common2}
\begin{equation}\label{eq3-5}
\begin{aligned}
 & c_1(t)= [r_B^2+r_A^2 p(t)] c_{10} - r_A r_B [1-p(t)] c_{20}, \\
 & c_2(t)= [r_A^2+r_B^2 p(t)] c_{20} - r_A r_B [1-p(t)] c_{10},
\end{aligned}
\end{equation}
and by denoting $\delta_A=\delta_B \equiv \delta$, $p(t)$ can be written as
\begin{equation}\label{eq3-6}
 p(t)= e^{-\frac{1}{2}(\lambda-i\delta)t}\left[\cosh\left(\frac{d t}{2}\right)+\frac{\lambda-i\delta}{d}\sinh\left(\frac{d t}{2}\right)\right],
\end{equation}
with $d= \sqrt{(\lambda-i\delta)^2-4R^2\lambda^2}$, and $R = \alpha_T W /\lambda$ is a dimensionless parameter. The strong and weak couplings, corresponding to the good and bad cavity limits, are described by $R\gg 1$ and $R \ll 1$, respectively \cite{common1}. Equation \eqref{eq3-5} indicates that for the equal detuning scenario, there is a subradiant (dark) state $|\psi_{-}\rangle= r_B |10\rangle - r_A |01\rangle$ which is decoherence-free.

For $\omega_A \neq \omega_B$ and the initial state in Eq. \eqref{eq3-3}, $c_1(t)$ and $c_2(t)$ can be found in Refs. \cite{common2,common0} (for brevity, we do not list them here). As is emphasized in Refs. \cite{common1,common2}, the solutions for the initial state $|\psi(0)\rangle$ are exact as neither the Born nor the Markov approximation was used.

\section{Preservation and generation of NAQI} \label{sec:4}
With the preliminaries in Secs. \ref{sec:2} and \ref{sec:3} at hand, we investigate dynamics of NAQI in this section, putting emphasis on the off-resonant preservation and generation of it. As the subradiant state $|\psi_{-}\rangle$ does not decay in time for the resonant case, we consider the initial state of Eq. \eqref{eq3-3} with $c_{10}=c_{20}=1/\sqrt{2}$ (referred to as $|\Psi^{+}\rangle$ in the following) in Sec. \ref{sec:4a}. Moreover, to show the possible off-resonant generation of NAQI, we will consider the initial product state $|10\rangle$ in Sec. \ref{sec:4b}, which corresponds to $c_{10}=1$ and $c_{20}=0$. For any $\rho_{AB}$ obtained from Eq. \eqref{eq3-4}, the method for calculating $\mathcal{N}_{\alpha}(\rho_{AB})$ ($\alpha \in\{tr,re,g\}$) is given in Appendix \ref{sec:B}. Moreover, to lighten notation, we will use the simplified notation $\delta= \delta_0$ to represent $\delta_A= \delta_B = \delta_0$. Of course, for the situation in which $\delta_A \neq \delta_B$, we will specify their values separately.

\subsection{NAQI preservation} \label{sec:4a}
\begin{figure}
\centering
\resizebox{0.45 \textwidth}{!}{%
\includegraphics{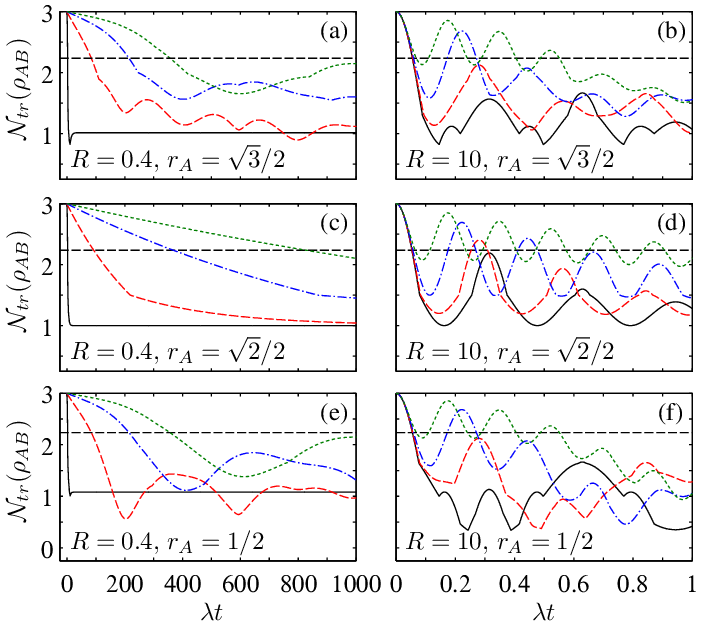}}
\caption{$\mathcal{N}_{tr}(\rho_{AB})$ versus $\lambda t$ for the initial state $|\Psi^{+}\rangle$ with different $R$ and $r_A$, where the solid black, dashed red, dash-dotted blue, and dotted green lines correspond to $\delta=0$, $10\lambda$, $20\lambda$, and $30\lambda$, respectively. The horizontal dashed lines show the bound $\mathcal{I}_{tr,\mathcal{B}}$ larger than which there is trace norm of NAQI in $\rho_{AB}$.} \label{fig:2}
\end{figure}

We first consider the preservation of NAQI with $\delta_A = \delta_B$. In Fig. \ref{fig:2} we show $\mathcal{N}_{tr}(\rho_{AB})$ as a function of $\lambda t$ with different $R$ and $r_A$, and as pointed out in Sec. \ref{sec:2}, it takes the maximum of 3 initially. In the bad cavity limit, as exemplified in the left three panels of this figure, $\mathcal{N}_{tr}(\rho_{AB})$ decays rapidly for the resonant case (i.e., $\delta= 0$), and the trace norm of NAQI disappears after a very short evolution time $t_c$, e.g., $\lambda t_c \approx 1.7748$, $1.7011$, and $1.7748$ for $R$ and $r_A$ given in Figs. \ref{fig:2}(a), \ref{fig:2}(c), and \ref{fig:2}(e), respectively. By increasing the frequency detuning $\delta$, the decay rate of $\mathcal{N}_{tr}(\rho_{AB})$ is significantly suppressed and the trace norm of NAQI exists in an obviously extended time region. One can also note that $\mathcal{N}_{tr}(\rho_{AB})$ with different $r_A$, which describes the relative coupling strength of the qubits to the common reservoir, show different dynamical patterns. Specifically, $\mathcal{N}_{tr}(\rho_{AB})$ progressively decreases as time evolves for the equal coupling scenario (i.e., $r_A=\sqrt{2}/2$) and shows damped oscillations for the unequal coupling scenario. Moreover, in the presence of frequency detuning (i.e., $\delta>0$), the survival time of the trace norm of NAQI for the equal coupling scenario in Fig. \ref{fig:2}(c) is obviously longer than that for the unequal coupling scenario in Figs. \ref{fig:2}(a) and \ref{fig:2}(e).

In the good cavity limit, as exemplified in the right panels of Fig. \ref{fig:2}, $\mathcal{N}_{tr}(\rho_{AB})$ for both the resonant and off-resonant cases show rapid oscillations with the evolving of time. For $\delta=0$, the trace norm of NAQI also disappears after a very short evolution time $t_c$ [e.g., $\lambda t_c \approx 0.0552$, $0.0534$, and $0.0552$ for $R$ and $r_A$ given in Figs. \ref{fig:2}(b), \ref{fig:2}(d), and \ref{fig:2}(f), respectively], and such a critical time is only slightly increased with the increasing $\delta$. But different from the bad cavity limit case, for large detunings $\mathcal{N}_{tr}(\rho_{AB})$ exhibits a damped oscillatory profile, enveloped by an exponential decay, and there are revivals for the trace norm NAQI [cf. the dash-dotted and dotted lines in Figs. \ref{fig:2}(b), \ref{fig:2}(d), and \ref{fig:2}(f)]. The revival regions can be greatly extended by increasing $\delta$, especially for the equal coupling scenario in Fig. \ref{fig:2}(d). The physical reason for such an enhancement can be attributed to the non-Markovian effect triggered by increasing $\delta$ (see Sec. \ref{sec:3}), which causes a backflow of information from the reservoir to the two-qubit system.

Nonetheless, as shown in Fig. \ref{fig:2}, it is evident that the survival time for the trace norm of NAQI in the bad cavity limit is much longer than that in the good cavity limit. Physically, this is because the coupling of the qubits is mediated by the cavity mode through exchanging virtual photons, and in the bad cavity limit the two qubits are only weakly coupled and the NAQI is less affected by the cavity losses. For the large off-resonant case $\delta_{A,B} \gg R\lambda \equiv \mathcal{R}$ (see the dotted lines in Fig. \ref{fig:2}), with $\mathcal{R}$ being the vacuum Rabi frequency, the system dynamics can be effectively characterized by a dispersive Hamiltonian $H_{\mathrm{eff}}$ describing the coupling of the qubits with a single-mode cavity field, given by \cite{common2}
\begin{equation} \label{eq4-1}
 H_{\mathrm{eff}}= \sum_{n=A,B} \left[\frac{\mathcal{R}^2r_n^2}{\delta_n}\sigma^n_{+}\sigma^n_{-}
                   +\frac{\mathcal{R}^2 r_A r_B}{2\delta_n} \big(\sigma^A_{+}\sigma^B_{-} + \sigma^A_{-}\sigma^B_{+} \big)\right],
\end{equation}
where the first part describes the Stark shifts due to the dispersive interaction of the qubits with the cavity vacuum, and the second part describes their dipole-dipole coupling induced by the cavity mode. Equation \eqref{eq4-1} shows that both the Stark shifts and the dipole-dipole interaction are proportional to $\mathcal{R}^2 /\delta_{A,B}$; thereby in the dispersive regime, the decay rate is noticeably suppressed and the NAQI is effectively preserved both in the bad and good cavity limits. For the equal detuning case with $r_A^2= 1/2$ and $\delta \gg \lambda,\mathcal{R}$, following the similar steps as in Ref. \cite{common2}, one could show that $p(t)\approx e^{-\mathcal{R}^2 (\lambda+i\delta) t/\delta^2}$ and $\mathcal{N}_{tr}(\rho_{AB}) \approx 3e^{-2(\mathcal{R}/\delta)^2\lambda t}$. Since in this case $(\mathcal{R}/\delta)^2$ is very small, the decay of $\mathcal{N}_{tr}(\rho_{AB})$ is slowed down noticeably and the NAQI is well preserved compared to the resonant case. In fact, the cavity photon in this case is only virtually excited \cite{common2}; thereby, the cavity losses do not significantly affect the dynamical patterns of $\mathcal{N}_{tr}(\rho_{AB})$. Moreover, note that Eq. \eqref{eq4-1} holds for $\delta_{A,B} \gg \mathcal{R}$. For the finite $\delta_{A,B}$, the approximate $\rho_{AB}^{\mathrm{ap}}(t)$ will deviate from $\rho_{AB}(t)$ obtained from Eq. \eqref{eq3-4}. For example, when $r_A^2=0.5$, one can show that for $R=0.4$, the minimum fidelity $\min_{t}F[\rho_{AB}(t), \rho_{AB}^{\mathrm{ap}}(t)]$ increases from $0.9985$ to $0.9999$ when $\delta/\mathcal{R}$ increases from $20$ to $100$, and for $R=1$, it increases from $0.9974$ to $0.9998$ when $\delta/\mathcal{R}$ increases from $20$ to $100$. Even for $R=10$, one still has $\min_{t}F[\rho_{AB}(t), \rho_{AB}^{\mathrm{ap}}(t)]>0.9998$ for $\delta/\mathcal{R}=100$. This shows that Eq. \eqref{eq4-1} can provide a reliable description of the system dynamics when $\delta/\mathcal{R}>100$.

\begin{figure}
\centering
\resizebox{0.45 \textwidth}{!}{%
\includegraphics{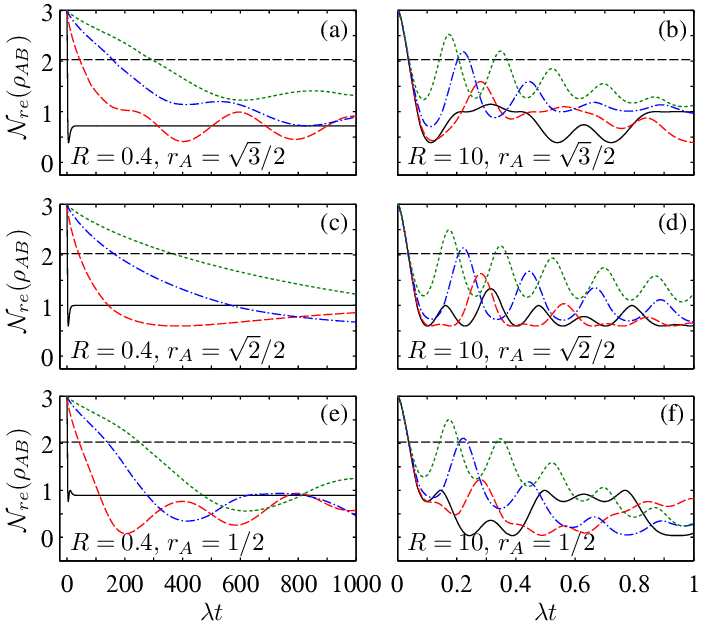}}
\caption{$\mathcal{N}_{re}(\rho_{AB})$ versus $\lambda t$ for the initial state $|\Psi^{+}\rangle$ with different $R$ and $r_A$, where the solid black, dashed red, dash-dotted blue, and dotted green lines correspond to $\delta=0$, $10\lambda$, $20\lambda$, and $30\lambda$, respectively. The horizontal dashed lines show the bound $\mathcal{I}_{re,\mathcal{B}}$ larger than which there is relative entropy of NAQI in $\rho_{AB}$.} \label{fig:3}
\end{figure}

In Fig. \ref{fig:3} we show the time evolution of the relative entropy of $\mathcal{N}_{re}(\rho_{AB})$ with different $R$ and $r_A$. It can be seen that it shows structurally similar dynamical patterns to that of $\mathcal{N}_{tr}(\rho_{AB})$, that is, in the bad cavity limit, the survival time of the relative entropy of NAQI increases with the increasing frequency detuning $\delta$, while in the good cavity limit, it also displays damped oscillations and there are revivals during certain time regions; meanwhile, for the large detuning case $\mathcal{N}_{re}(\rho_{AB})$ also exhibits an exponential decay envelope, see the dotted lines in the right panels of Fig. \ref{fig:3}. Despite these similarities, by carefully comparing Figs. \ref{fig:2} and \ref{fig:3} one can note that there are several small differences between $\mathcal{N}_{tr}(\rho_{AB})$ and $\mathcal{N}_{re}(\rho_{AB})$; for example, the survival time for the relative entropy of NAQI is significantly shorter than that for the trace norm of NAQI, and this indicates that the former is less efficient than the latter in revealing the NAQI. Moreover, while the survival time for the trace norm of NAQI is the same for $r_A =\sqrt{3}/2$ and $1/2$, this is not the case for the relative entropy of NAQI. The very reason behind this difference may be their respective definitions, as the former is defined from a geometric perspective and it depends on the imaginarity parts of $\rho_{B|\Pi_i^a}$ in the MUBs, while the latter is defined from an entropic perspective and it depends on all the elements of  of $\rho_{B|\Pi_i^a}$ in the MUBs [see Eq. \eqref{eq2-2}].

\begin{figure}
\centering
\resizebox{0.45 \textwidth}{!}{%
\includegraphics{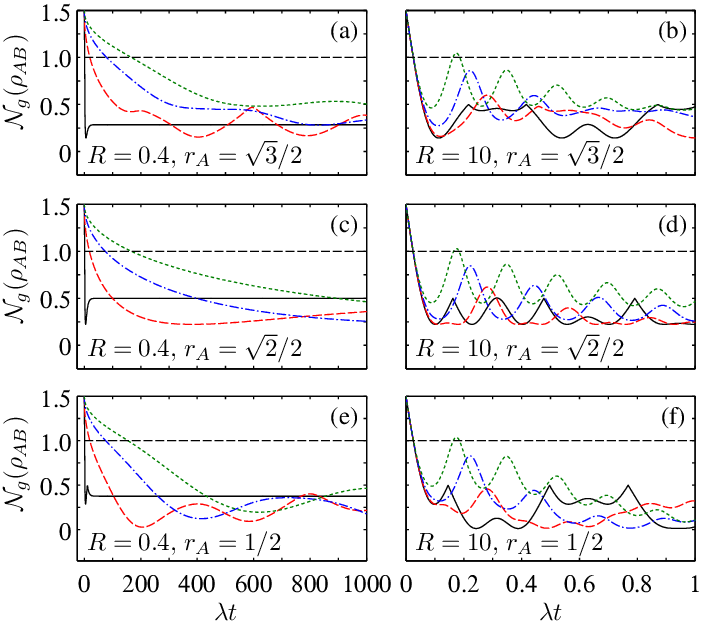}}
\caption{$\mathcal{N}_g(\rho_{AB})$ versus $\lambda t$ for the initial state $|\Psi^{+}\rangle$ with different $R$ and $r_A$, where the solid black, dashed red, dash-dotted blue, and dotted green lines correspond to $\delta=0$, $10\lambda$, $20\lambda$, and $30\lambda$, respectively. The horizontal dashed lines show the bound $\mathcal{I}_{g, \mathcal{B}}$ larger than which there is geometric NAQI in $\rho_{AB}$.} \label{fig:4}
\end{figure}

As for $\mathcal{N}_g(\rho_{AB})$, its time evolution under different parameters $R$ and $r_A$ is shown in Fig. \ref{fig:4}. Consistent with those in Figs. \ref{fig:2} and \ref{fig:3}, in the bad cavity limit, the survival time of the geometric NAQI can also be significantly increased by increasing the detuning $\delta$, while in the good cavity limit, there are revivals of the geometric NAQI only for large enough $\delta$. Moreover, the survival time regions for the NAQI in this case are further shortened compared to the previous two cases. This indicates that the fidelity-based geometric measure is also not very efficient in capturing the NAQI, which follows from the similar reason as the difference between $\mathcal{N}_{tr}(\rho_{AB})$ and $\mathcal{N}_{re}(\rho_{AB})$. Thereby, we focus in the following only on the trace norm of NAQI. As a matter of fact, our calculation reveals that when one chooses other system parameters (e.g., the parameters $\delta_A \neq \delta_B$) or other initial states (see, e.g., the initial state $|10\rangle$ analyzed in Sec. \ref{sec:4b}), the behaviors of both $\mathcal{N}_{re}(\rho_{AB})$ and $\mathcal{N}_g(\rho_{AB})$ are also structurally similar to that of $\mathcal{N}_{tr}(\rho_{AB})$ and no new physics can be obtained. This is also a reason that we take $\mathcal{N}_{tr}(\rho_{AB})$ as the main quantity afterward.

\begin{figure}
\centering
\resizebox{0.45 \textwidth}{!}{%
\includegraphics{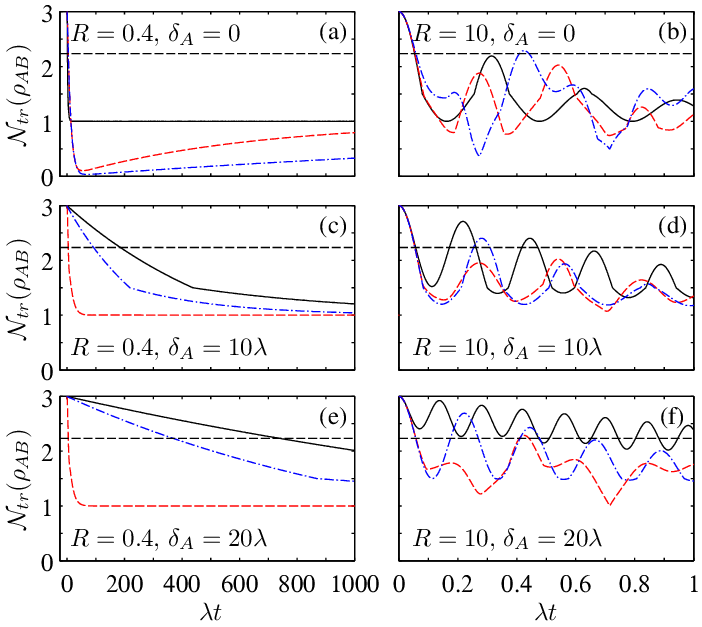}}
\caption{$\mathcal{N}_{tr}(\rho_{AB})$ versus $\lambda t$ for the initial state $|\Psi^{+}\rangle$ with $r_A= \sqrt{2}/2$ and different $R$, $\delta_A$, and $\delta_B$. The solid black, dashed red, and dash-dotted blue lines in (a) and (b) [(c)--(f)] are plotted for $\delta_B=0$, $10\lambda$, and $20\lambda$ [$\delta_B=-\delta_A$, $0$, and $\delta_A$], respectively.} \label{fig:5}
\end{figure}

While Figs. \ref{fig:2}--\ref{fig:4} illustrate the situation of equal detunings, in Fig. \ref{fig:5} we show time evolution of $\mathcal{N}_{tr}(\rho_{AB})$ for a more general situation in which the detunings of the two qubits from the main cavity frequency are different and the two qubits are equally coupled to the reservoir. Our purpose is to see whether the unequal detunings can help to prolong the survival time of the NAQI. Since $\mathcal{N}_{tr}(\rho_{AB})$ is invariant under the substitution $(\delta_A,\delta_B) \rightarrow (-\delta_A,-\delta_B)$, we focus only on the case of $\delta_A \geqslant 0$. For $\delta_A=0$, the survival region of NAQI can be extended very \emph{}slightly by increasing $|\delta_B|$, and there are revivals of NAQI in the good cavity limit only for large enough $|\delta_B|$ [cf. the dash-dotted blue line in Fig. \ref{fig:5}(b)]. For $\delta_A>0$, it is evident that a negative $\delta_B$ is more efficient than that of a vanishing or a positive one in extending the survival region of NAQI. This shows that if the frequencies of the two qubits are symmetrically detuned from the fundamental frequency of the cavity field (i.e., $\delta_B = -\delta_A$), the NAQI can be more efficiently preserved than that under the same frequency detunings.

Moreover, in Fig. \ref{fig:5} we considered only the equal coupling scenario. For the unequal coupling scenario, a similar behavior can also be observed, and the reason we do not present the plots here is that in the dispersive regime (i.e., $|\delta_A|, |\delta_B| \gg \mathcal{R}$), the main features of the system can be effectively described by $H_{\mathrm{eff}}$ in Eq. \eqref{eq4-1}, from which one can obtain that for $\delta_B = -\delta_A$, the dipole-dipole coupling term vanishes. Thus the Stark shifts term dominates and the effective decay rate is determined by the total coupling parameter $\alpha_T$ (note that $\mathcal{R} \propto \alpha_T$) rather than the relative coupling parameter $r_A$. Thereby, $\mathcal{N}_{tr}(\rho_{AB})$ will be insensitive to the variation of $r_A$.

\subsection{NAQI generation} \label{sec:4b}
\begin{figure}
\centering
\resizebox{0.45 \textwidth}{!}{%
\includegraphics{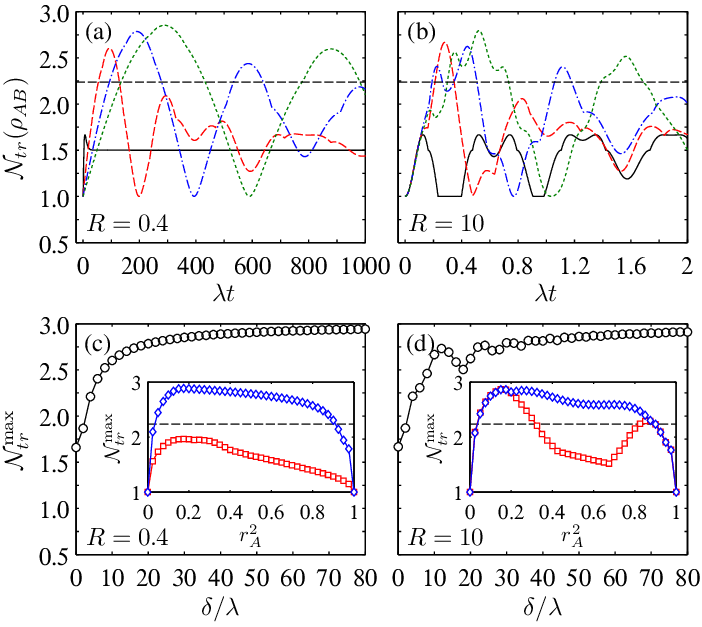}}
\caption{(Top) $\mathcal{N}_{tr}(\rho_{AB})$ versus $\lambda t$ for the initial state $|10\rangle$ with $r_A= \sqrt{2}/2$ and different $R$, where the solid black, dashed red, dash-dotted blue, and dotted green lines correspond to $\delta=0$, $10\lambda$, $20\lambda$, and $30\lambda$, respectively. (Bottom) The maximum $\mathcal{N}_{tr}^{\max}$ generated from $|10\rangle$ with $r_A=\sqrt{2}/2$ and different $R$. Insets: $\mathcal{N}_{tr}^{\max}$ versus $r_A^2$ for $\delta =0$ (squares) and $20\lambda$ (diamonds), respectively.} \label{fig:6}
\end{figure}

In this section we focus on the generation of NAQI from the initial product state $|10\rangle$, for which there is no NAQI initially. In Figs. \ref{fig:6}(a) and \ref{fig:6}(b) we show $\mathcal{N}_{tr}(\rho_{AB})$ as a function of $\lambda t$ with different $R$, where the two qubits are equally coupled to the reservoir. For the resonant case (i.e., $\delta=0$), it can be seen that there is no NAQI being generated. The presence of detunings enhances $\mathcal{N}_{tr}(\rho_{AB})$, and when $\delta$ is large enough, there will be NAQI in certain time regions. These regions extend with the increasing $\delta$ both in the bad and in the good cavity limits, but the detunings in the good cavity limit induce the generation of NAQI at short times compared to that in the bad cavity limit. The generated maximum $\mathcal{N}_{tr}(\rho_{AB})$ (referred to as $\mathcal{N}_{tr}^{\max}$ for brevity) can be enhanced by increasing the detuning $\delta$. In the bad cavity limit, as is illustrated in Fig. \ref{fig:6}(c), $\mathcal{N}_{tr}^{\max}$ increases with the increase of $\delta$ and approaches 3 for large enough $\delta$, e.g., it becomes larger than $\mathcal{I}_{tr,\mathcal{B}}$ when $\delta \gtrsim 4.2051\lambda$ and takes the value of about $2.9424$ for $\delta= 80\lambda$. In the good cavity limit, as evident from Fig. \ref{fig:6}(d), although $\mathcal{N}_{tr}^{\max}$ shows several weak oscillations (e.g., there is a pronounced cusp at which $\delta \approx 18.7287\lambda$ and $\mathcal{N}_{tr}^{\max} \approx 2.4698$), it becomes larger than $\mathcal{I}_{tr,\mathcal{B}}$ for $\delta \gtrsim 4.1072 \lambda$ and a further calculation shows that in theory, $\mathcal{N}_{tr}^{\max}$ also approaches $3$ for large enough $\delta$, e.g., $\mathcal{N}_{tr}^{\max} \approx 2.9116$ at $\delta= 80\lambda$. One can also note that $\mathcal{N}_{tr}^{\max}$ generated at $\delta=80\lambda$ is slightly higher in the bad cavity limit than in the good cavity limit. The mechanism underlying this is the physical effect tied to how ``deep" the system is in the dispersive regime, characterized by $\delta \gg \mathcal{R}$. Specifically, in the bad cavity limit $R = 0.4$, the ratio $\delta/\mathcal{R}= 200$, heavily suppressing real cavity excitation and subsequent decoherence. In the good cavity limit $R = 10$, however, $\delta/\mathcal{R}= 8$, suggesting that the system is only marginally dispersive and still suffers some real interaction with the lossy cavity mode.

The coupling scenario of the two qubits with the reservoir also affects $\mathcal{N}_{tr}^{\max}$ significantly. As shown in the insets of Fig. \ref{fig:6}, $\mathcal{N}_{tr}^{\max}$ does not take its maximum at $r_A^2=0.5$ (i.e., the equal coupling case). In the bad cavity limit, there is no NAQI being generated for the resonant case, while for the off-resonant case denoted by the blue diamonds, $\mathcal{N}_{tr}^{\max}$ reaches its peak value of about $2.8813$ at $r_A^2 \approx 0.1721$, and there is a wide $r_A^2$ region in which there is NAQI. In the good cavity limit, however, values of $\mathcal{N}_{tr}^{\max}$ close to 3 can be generated for both the resonant and off-resonant situations. Specifically, for the parameters in Fig. \ref{fig:6}(d), if $\delta=0$ ($\delta=20\lambda$), $\mathcal{N}_{tr}^{\max}$ reaches its peak value of about $2.8694$ ($2.8664$) at $r_A^2 \approx 0.1561$ ($0.1623$). Nonetheless, the $r_A^2$ region during which there is NAQI for the off-resonant situation is still wider than that for the resonant situation. The generation of a nearly optimal NAQI from $|10\rangle$ can be explained as follows. From Eq. \eqref{eq3-6} it can be shown that when $r_A^2= 0.5$, $\delta \gg \mathcal{R}$, and $\delta \gg \lambda$, $p(t_c)\approx -i$ at $t_c\approx \pi\delta/ 2\mathcal{R}^2$. This, together with Eq. \eqref{eq3-5}, yields $c_1(t_c) \approx (1-i)/2$ and $c_2(t_c) \approx - (1+i)/2$. Thereby, $\rho_{AB}(t_c) \rightarrow |\psi\rangle\langle\psi|$, with $|\psi\rangle= (|10\rangle-i|01\rangle)/\sqrt{2}$ being a maximally entangled state, hence $\mathcal{N}_{tr}^{\max} \approx 3$. Note that here the maximum NAQI generated from $|10\rangle$ occurs at symmetric coupling as $\delta \gg \mathcal{R}$. For the finite $\delta$ case, $\mathcal{N}_{tr}^{\max}$ takes its maximum at asymmetric coupling ratios $r_A^2$. Physically, this phenomenon has its roots in the asymmetric property of the initial state $|10\rangle$, for which the first qubit is excited and the second one is in the ground state. Thereby, a relatively weak coupling of the first qubit with the reservoir is favorable for generating the maximum NAQI, as the effective qubit-qubit interaction is mediated by the cavity mode through exchanging virtual cavity photons. Moreover, for the resonant case, while $\mathcal{N}_{tr}^{\max}$ as a function of $r_A^2$ in the bad cavity limit exhibits only a single peak, it exhibits two peaks in the good cavity limit. We make a heuristic analysis of this observation by considering the extreme points of the nonzero elements of $\rho_{AB}(t)$. As $p\in \mathbb{R}$ for $\delta=0$, one can show that the extreme points for $c_1^2$ and $c_2^2$ either give to a diagonal $\rho_{AB}$ or are covered by that of $c_1 c_2$. For $c_1c_2$, by considering its derivative with respect to $r_A$, one can find that  the extreme points occur at
\begin{equation} \label{eq-n1}
 r_A^2= \frac{5-3p\pm \sqrt{9p^2-14p+9}}{8(1-p)},
\end{equation}
and thus one has
\begin{equation} \label{eq-n2}
 r_A^2 \in
  \left\{
   \begin{aligned}
    & [0.25,0.5)                      & \text{if}~ p \geqslant 0, \\
    & (\mu_{-},0.25) \cup (\mu_{+},1) & \text{if}~ p < 0,
   \end{aligned}
  \right.
\end{equation}
where $\mu_\pm =1/2\pm \sqrt{2}/4$. If $R \leqslant 1/2$, one always has $p(t) \geqslant 0$, and $c_1 c_2$ has only one extreme point; thus it is understandable that the resulting $\mathcal{N}_{tr}^{\max}$ exhibits a single peak, see the inset of Fig. \ref{fig:6}(c). If $R>1/2$, however, $p(t)$ can be negative in certain time regions. Then Eq. \eqref{eq-n2} implies that it is possible that $c_1 c_2$ has two extreme points, which yields the two peaks of $\mathcal{N}_{tr}^{\max}$ in the good cavity limit [see the inset of Fig. \ref{fig:6}(d)].

For the initial product state $|01\rangle$, $\mathcal{N}_{tr}(\rho_{AB})$ can be obtained directly from that for $|10\rangle$ under the substitution $r_A^2 \rightarrow 1-r_A^2$, thereby a high degree of NAQI can also be generated both in the bad and in the good cavity limits.

While we considered in the above the case of equal detunings, in the light of the observation for the NAQI preservation in Sec. \ref{sec:4a}, it is natural to ask whether the case of symmetric detunings $\delta_B=-\delta_A$ is more efficient in generating the NAQI. To answer this question we performed numerical calculation and it was found that the answer to this question is negative, that is, the NAQI generated from $|10\rangle$ under $\delta_B=-\delta_A$ is much smaller than that under $\delta_B=\delta_A$. This behavior is in sharp contrast to that in Sec. \ref{sec:4a}, where it was shown that a symmetric detuning is more efficient in preserving the NAQI. As $\mathcal{N}_{tr}(\rho_{AB})$ is not analytically solvable, it is hard to explain this contrast analytically for the general case. But in the dispersive regime (i.e., $\delta_{A,B} \gg \mathcal {R}$), the system dynamics can be analyzed by considering the effective Hamiltonian \eqref{eq4-1} and the cavity losses. Specifically, for $\delta_B=-\delta_A$, the dipole-dipole coupling term in $H_{\mathrm{eff}}$ vanishes and the remaining Stark shifts term cannot generate correlations from $|10\rangle$; this explains why it is inefficient at generating NAQI. For the state $|\Psi^{+}\rangle$, the NAQI takes its maximum initially, and for $\delta_B=-\delta_A$, the only effect is the NAQI decay induced by the cavity losses. For $\delta_B = \delta_A$, however, the cavity losses induce decay of the NAQI and the effective dipole-dipole coupling leads to its oscillatory behavior. Then the combined effects yield a damped oscillatory profile of $\mathcal{N}_{tr}(\rho_{AB})$, enveloped by an exponential decay.

Moreover, one may also have concerns about the stationary NAQI in the infinite-time limit $t \rightarrow \infty$, for which it follows from Eq. \eqref{eq3-5} that $c_1(\infty)=r_B \beta_{-}$ and $c_2(\infty)= -r_A \beta_{-}$, where $\beta_{-}= r_B c_{10}-r_A c_{20}$. Then one can obtain numerically that for the initial state $|\Psi^{+}\rangle$, $\mathcal{N}_{tr}(\rho_{AB}(\infty))$ takes its maximum of about $1.2181$ at $r_A \approx 0.2356$, while for the initial state $|10\rangle$, it takes its maximum of about $1.9651$ at $r_A \approx 0.4185$. Thus it can be seen that there is no stationary NAQI for both of the two initial states.

Before ending this section, we remark that although we discussed $R \!\sim\! 10$ and $\delta_n/\lambda \!\sim\! 10^2$ ($n=A,B$), one still has $W \ll \omega_n$ and $\delta_n\ll \omega_n$. This justifies the RWA used in Eq. \eqref{eq3-1} \cite{common1,common2}. In fact, $\omega_n$ can be tuned to the order of GHz for superconducting qubits in cavity QED setups \cite{nature1,nature2,nature3}, and $\delta_n$ can be tuned by tuning $\omega_n \approx \sqrt{8E_J E_C}/\hbar$, with $E_J$ ($E_C$) being the Josephson (charging) energy \cite{nature3}. The qubit-field coupling can be tuned by tuning the position of the qubits in the cavity-field standing wave, and $R \sim 10$ is experimentally achievable \cite{nature1,nature2,nature3,apl}.

\section{Preservation and generation Of DIA} \label{sec:5}
In this section we consider the off-resonant protection and generation of DIA. Different from NAQI, it deals with the task in which Alice aims to assist Bob to locate an optimal amount of imaginarity on  $B$ via LQRCC, see Fig. \ref{fig:1}(b). For the initial state $|\psi(0)\rangle$ of Eq. \eqref{eq3-3}, $v_{02}=v_{32}=0$, $v_{12}= 2\mathrm{Im}[c_1(t)c_2^*(t)]$, and $v_{22}=2\mathrm{Re}[c_1(t)c_2^*(t)]$, with $\mathrm{Im}[\cdot]$ and $\mathrm{Re}[\cdot]$ being the imaginary and real parts of a number, respectively. Then it follows from Eq. \eqref{eq2-9} that
\begin{equation} \label{eq5-1}
 F_a(\rho_{AB})= \frac{1}{2}+|c_1(t)c_2^*(t)|,
\end{equation}
and in alignment with Sec. \ref{sec:4}, we still consider the initial Bell state $|\Psi^+\rangle$ and product state $|10\rangle$. Since $v_{02}=0$ for this case, then as mentioned earlier, the assisted imaginarity distillation has quantum advantage when $F_a(\rho_{AB})> 1/2$.

\subsection{DIA preservation} \label{sec:5a}
\begin{figure}
\centering
\resizebox{0.45 \textwidth}{!}{%
\includegraphics{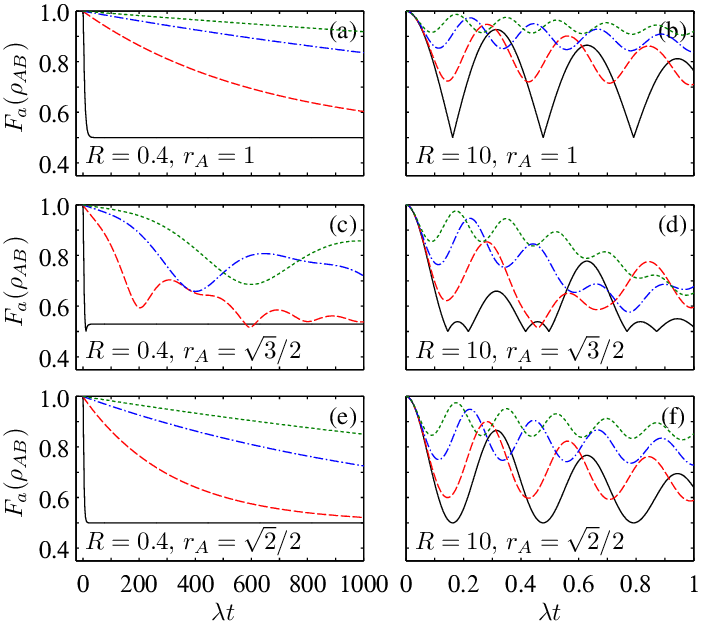}}
\caption{$F_a(\rho_{AB})$ versus $\lambda t$ for the initial state $|\Psi^{+}\rangle$ with different $R$ and $r_A$, where the solid black, dashed red, dash-dotted blue, and dotted green lines correspond to $\delta=0$, $10\lambda$, $20\lambda$, and $30\lambda$, respectively.} \label{fig:7}
\end{figure}

For $|\Psi^+\rangle$, one has $c_{1,2}(t)=[1+(r_{A,B}^2+r_A r_B)(p-1)]/\sqrt{2}$ when $\omega_A=\omega_B$, from which $F_a(\rho_{AB})$ can be obtained numerically; in particular, it is symmetric with respect to $r_A^2=0.5$. In Fig. \ref{fig:7}, $F_a(\rho_{AB})$ is plotted against $\lambda t$ with different $R$ and $r_A$. In the bad cavity limit, it is evident that $F_a(\rho_{AB})$ decays quickly for the resonant case, and although there is a very small cusp for $r_A^2 \neq 1$ and $0.5$, $F_a(\rho_{AB})$ always approaches an asymptotic value determined by $r_A$. A further calculation shows that when $r_A^2$ increases from $0.5$ to $1$, this asymptotic value first increases from $0.5$ to a maximum of about $0.5625$ at $r_A^2 \approx 0.9330$ and then decreases gradually to $0.5$ again. For the off-resonant situation, if $r_A^2=1$ or $0.5$, $F_a(\rho_{AB})$ also progressively decreases with the evolving time, and its magnitude increases with the increasing $\delta$ [cf. Figs. \ref{fig:7}(a) and \ref{fig:7}(e)]. If $r_A^2 \in (0.5, 1)$, as evident from Fig. \ref{fig:7}(c), damped oscillations appear in $F_a(\rho_{AB})$ and its value is somewhat decreased compared to the cases of $r_A^2=1$ and $0.5$. Physically, the mechanism underlying the enhancement of DIA is the same as that for the enhanced NAQI in Fig. \ref{fig:2}. For $\delta \gg \lambda \gg \mathcal{R}$ and $r_A= \sqrt{2}/2$, following the similar steps as in Ref. \cite{common2}, one can obtain $F_a(\rho_{AB}) \approx e^{-(\mathcal{R}/\delta)^2\lambda t}$. As $(\mathcal{R}/\delta)^2 \ll 1$, it is understandable that the DIA can be well preserved in a very long $\lambda t$ region.

In the good cavity limit $R=10$, as illustrated in the right three panels of Fig. \ref{fig:7}, $F_a(\rho_{AB})$ shows more rich dynamics for any $r_A$; specifically, it shows rapid oscillations as a function of time due to the strong non-Markovianity, and the quantum advantage of assisted imaginarity distillation is well preserved in the short time region, especially for the case of large detunings. But in the long time region, $F_a(\rho_{AB})$ still decays to an asymptotic value which is the same as that in the bad cavity limit (for brevity, we do not show the plots here). This implies that while in the short time region the non-Markovianity dominates, in the long time region its effect will be weakened and the detrimental effect turn to dominates.

\begin{figure}
\centering
\resizebox{0.45 \textwidth}{!}{%
\includegraphics{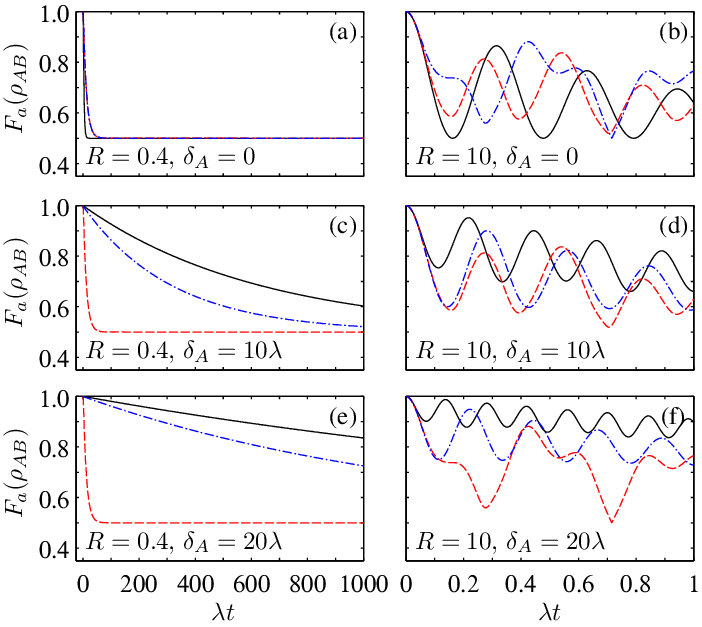}}
\caption{$F_a(\rho_{AB})$ versus $\lambda t$ for the initial state $|\Psi^{+}\rangle$ with $r_A=\sqrt{2}/2$ and different $R$ and $\delta_{A,B}$. The solid black, dashed red, and dash-dotted blue lines in (a) and (b) [(c)--(f)] are plotted for $\delta_B=0$, $10\lambda$, and $20\lambda$ [$\delta_B=-\delta_A$, $0$, and $\delta_A$], respectively.} \label{fig:8}
\end{figure}

Figure \ref{fig:8} illustrates the $\lambda t$ dependence of $F_a(\rho_{AB})$ for which the two qubits have unequal detunings, and due to the same reason stated in Sec. \ref{sec:4a}, we also fixed $r_A=\sqrt{2}/2$. In analogy to $\mathcal{N}_{tr}(\rho_{AB})$, $F_a(\rho_{AB})$ is also invariant under the substitution $(\delta_A, \delta_B) \rightarrow(-\delta_A, -\delta_B)$, and thus we take $\delta_A \geqslant 0$. For $\delta_A=0$, it can be seen that $F_a(\rho_{AB})$ decays to the asymptotic value of $0.5$ and it is insensitive to the increasing $\delta_B$ in the bad cavity limit. In the good cavity limit, although $F_a(\rho_{AB})$ oscillates quickly, the increasing $\delta_B$ still cannot help to effectively preserve the DIA. For $\delta_A \neq 0$, consistent with the NAQI in Fig. \ref{fig:5}, $F_a(\rho_{AB})$ is more efficiently preserved if $\delta_B$ takes an opposite sign to that of $\delta_A$, that is, a symmetric detuning of the qubits works better than the same detunings in preserving the DIA.

\subsection{DIA generation} \label{sec:5b}
\begin{figure}
\centering
\resizebox{0.45 \textwidth}{!}{%
\includegraphics{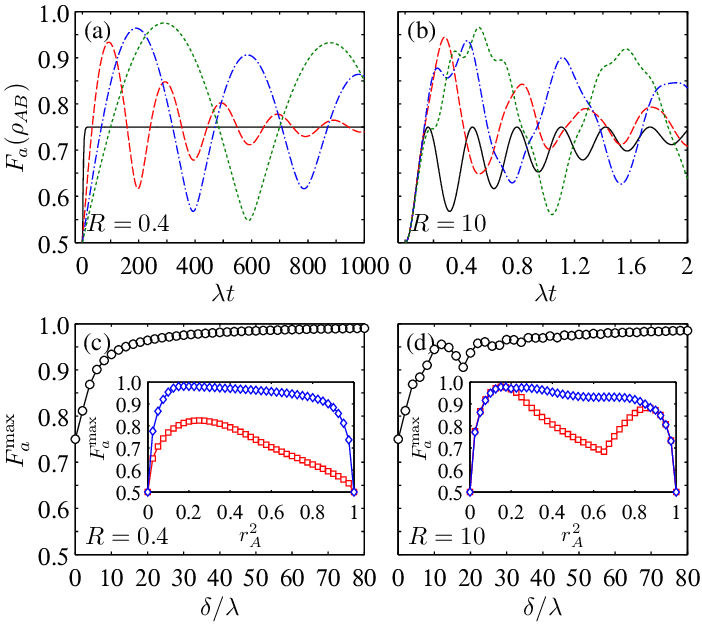}}
\caption{(Top) $F_a(\rho_{AB})$ versus $\lambda t$ for the initial state $|10\rangle$ with $r_A=\sqrt{2}/2$ and different $R$, where the solid black, dashed red, dash-dotted blue, and dotted green lines correspond to $\delta=0$, $10\lambda$, $20\lambda$, and $30\lambda$, respectively. (Bottom) The maximum $F_a^{\max}$ generated from $|10\rangle$ with $r_A=\sqrt{2}/2$ and different $R$. Insets: $F_a^{\max}$ versus $r_A^2$ for $\delta=0$ (squares) and $20\lambda$ (diamonds), respectively.} \label{fig:9}
\end{figure}

We now turn to consider the generation of DIA. As is shown in Figs. \ref{fig:9}(a) and \ref{fig:9}(b), for the initial state $|10\rangle$, $F_a(\rho_{AB})$ takes the value $0.5$, and the interaction with the cavity field generates a high degree of DIA both in the bad and in the good cavity limits. Of course, the time at which $F_a(\rho_{AB})$ reaches its maximum for the former scenario is much longer than that for the latter. Moreover, it can be seen from Fig. \ref{fig:9}(c) that in the bad cavity limit, the maximum $F_a(\rho_{AB})$ (referred to as $F_a^{\max}$) generated from $|10\rangle$ progressively increases with the increase of detuning, e.g., one has $F_a^{\max} > 0.98$ for $\delta \gtrsim 37.4639 \lambda$. In the good cavity limit, as is illustrated in Fig. \ref{fig:9}(d), although there are several weak oscillations for $F_a^{\max}$ (the most pronounced cusp corresponds to $\delta \approx 18.3256\lambda$ and $F_a^{\max} \approx 0.9013$), which is akin to $\mathcal{N}_{tr}^{\max}$ shown in Fig. \ref{fig:6}(d), $F_a^{\max}$ can also be enhanced by continuously enlarging $\delta$, e.g., in this case $F_a^{\max} > 0.98$ for $\delta \gtrsim 60.1045\lambda$. Additionally, in the large $\delta$ region, $F_a^{\max}$ is slightly larger in the bad cavity limit than in the good cavity limit. This is also due to the real effect tied to how ``deep" the system is in the dispersive regime (see Sec. \ref{sec:4b}).

As evident from the insets of Fig. \ref{fig:9}, the coupling scenario of the two qubits to the cavity field also plays an essential role in the DIA generation. In the bad cavity limit, it can be seen that $F_a^{\max}$ is not a monotonic function of $r_A^2$; specifically, it first increases from $0.5$ to a peak and then decreases gradually to $0.5$, e.g., the peak value is about $0.9802$ for $\delta=20\lambda$, which occurs at $r_A^2 \approx 0.1721$. In the good cavity limit, there is a pronounced decrease for $F_a^{\max}$ for the resonant scenario, and $F_a^{\max}$ reaches its peak value of about $0.9782$ at $r_A^2 \approx 0.1561$. For the off-resonant scenario, however, as is illustrated by the blue diamonds, the pattern of $F_a^{\max}$ is very similar to that in the bad cavity limit, and for $\delta= 20\lambda$, it reaches the peak value of about $0.9777$ at $r_A^2 \approx 0.1623$. Moreover, the peak value of $F_a^{\max}$ with respect to $r_A^2$ can be further enhanced by enlarging $\delta$, both in the bad and good cavity limits. Hence in theory, the DIA generated from $|10\rangle$ is almost perfect for large enough $\delta$. Having in mind that the evolved state approaches $|\psi\rangle= (|10\rangle-i|01\rangle)/\sqrt{2}$ at $t_c$ (see Sec. \ref{sec:4b}), it is immediate to understand the very reason behind this enhancement.
For the finite $\delta$, the physical reason for the observation that $F_a^{\max}$ generated from $|10\rangle$ occurs at the asymmetric coupling ratio is the same as that for the NAQI. Although for the general case, the complexity of the involved parameters in $F_a(\rho_{AB})$ makes it hard to give an analytical explanation, for the resonant case, as $p(t)\in \mathbb{R}$, from Eq. \eqref{eq5-1} one can show that the extreme points of $F_a(\rho_{AB})$ also locate in the range given in Eq. \eqref{eq-n2}. Of course, some of these extreme points may not be the maximum points. In fact, using $|c_1c_2|= r_A r_B (1-p) |r_B^2+r_A^2 p|$, one can check directly that the associated $r_A^2$ at which $F_a^{\max}$ reaches its maximum increases from $\mu_{-}\approx0.1464$ to $0.5$ when $p$ increases from $-1$ to 1. Since $p=1$ only when $t=0$ and it may be much larger than $-1$ for some system parameters, it is understandable that $F_a^{\max}$ does not take its maximum at $r_A^2= 0.5$ in general. As for the cusp at $r_A^2\approx 0.6509$ in the inset of Fig. \ref{fig:9}(d), it is an inflection point at which $\partial F_a^{\max}/\partial r_A \neq 0$. In fact, when $r_A^2$ is smaller (larger) than $0.6509$, $r_B^2+r_A^2 p(t_c)$ is positive (negative), where $t_c$ is the time when $F_a(\rho_{AB}(t_c))=F_a^{\max}$.

Equation \eqref{eq-n2} also explains why $F_a^{\max}$ exhibits only a single peak in the bad cavity limit. This is because for $R \leqslant 1/2$, one always has $p \geqslant 0$; hence the single peak lies in the range of $r_A^2 \in [0.25,0.5)$, see Fig. \ref{fig:9}(c). For $R > 1/2$, however, $p$ can be negative in certain time intervals, so in the good cavity limit, $F_a^{\max}$ as a function of $r_A^2$ in Fig. \ref{fig:9}(d) exhibits two peaks, which occur in $r_A^2 \in (\mu_{-},0.25)$ and $r_A^2 \in (\mu_{+},1)$, respectively.

Moreover, the DIA generated from $|10\rangle$ under $\delta_B=-\delta_A$ is much smaller than that generated under $\delta_B=\delta_A$, in analogy to the NAQI analyzed in Sec. \ref{sec:4b}, and the reason for such an inefficiency is similar to that for generating NAQI. So we do not show the plots here for the conciseness of this paper.

Finally, for the infinite-time case, it follows from Eqs. \eqref{eq3-5} and \eqref{eq5-1} that
\begin{equation} \label{eq5-2}
 F_a(\rho_{AB}(\infty))= \frac{1}{2}+r_A r_B |r_B c_{10}-r_A c_{20} |^2.
\end{equation}
Then for the initial state $|\Psi^{+}\rangle$, one can show that $F_a(\rho_{AB}(\infty))$ attains the maximum $9/16$ at $r_A^2 = (2 \pm \sqrt{3})/4$, while for $|10\rangle$, it attains the maximum $1/2+3\sqrt{3}/16$ at $r_A^2=1/4$. This indicates that even in the infinite time limit, one can still achieve a high degree of DIA preservation and generation.

\section{Summary and discussion} \label{sec:6}
In summary, we have investigated NAQI and DIA. For the former, two assisting players collaborate to show nonlocal advantage of steered imaginarity under MUBs, and for the latter, they collaborate to localize maximal imaginarity onto a target system. We focused on the case that the two qubits are coupled to a common lossy cavity and being prepared in the initial states with one excitation for which the system dynamics is solvable without performing the Born-Markov approximation. Based on this general solution, we analyzed in detail dynamics of the NAQI and DIA, aimed at revealing the role played on their preservation and generation by the experimentally adjustable parameters such as the relative coupling between the qubits and the cavity and the frequency detunings of the qubits from the cavity field. For the initial Bell state $|\Psi^+\rangle$, it was found that if the two qubits are equally coupled to the cavity, then in the bad cavity limit, the NAQI disappears permanently after a finite time, while the DIA decreases asymptotically. In the good cavity limit, however, the strong non-Markovianity induces NAQI revivals in a finite-time region and the DIA oscillates quite a few times before reaching an asymptotic value. Both in the bad and good cavity limits, the NAQI and DIA can be further enhanced by enlarging the detuning to large enough values, being comparable to their initial maxima; in particular, a symmetric detuning facilitates more efficient preservation of the NAQI and DIA than that of the same detunings.

We have also analyzed in detail generation of the NAQI and DIA both in the bad and good cavity limits. It was found that for the initial state $|10\rangle$, the interaction with the cavity field results in a high degree of NAQI and DIA generation at certain evolution times when the detunings of the two qubits are the same, and this is quite different from what we have seen for the preservation of these two resources. Moreover, the enlarged detunings can lead to the generation of greater NAQI and DIA, which is consistent with their preservation process. Additionally, compared to the equal coupling scenario, we can generate a higher degree of NAQI and DIA if the two qubits are unequally coupled to the cavity field, and the corresponding optimal relative coupling parameter $r_A^2$ depends on the dimensionless parameter $R$ and the detuning $\delta$.

As for the proof-of-principle experiments demonstrating the NAQI and assisted imaginarity distillation tasks, one can consider the all-optical setup \cite{ima_cp}, the cavity QED or circuit QED setting \cite{nature1,nature2,nature3,apl}, and the superconducting processors \cite{newref1}. For the prepared two-qubit state, the associated optimal measurements can be obtained numerically, and the experimental overhead of extracting the two quantities are different: while they both require general measurements on qubit $A$, the former one involves general rotations of qubit $B$ about different axes, and the latter one involves only real operations on $B$, which could be implemented more economically than the general complex operations in real experiments \cite{ima_prl,ima_pra}.

While we focused on the initial two-qubit state of Eq. \eqref{eq3-3}, the investigations can be generalized immediately to other initial states. For example, for the scenario of equal coupling and resonant frequencies, the evolution of the system for a general initial state can be solved via the pseudomode approach \cite{pmode1,pmode2,pmode3}, where the pseudomodes are auxiliary variables defined from the spectrum of the field. For $J(\omega)$ of Eq. \eqref{eq3-2}, the pseudomode master equation is given by \cite{pmode1,pmode2,pmode3,pmode4,pmode5,common3}
\begin{equation}\label{eq6-1}
\frac{\partial \tilde{\rho}}{\partial t}= -i[V, \tilde{\rho}] +\lambda (2 b\tilde{\rho} b^\dagger- b^\dagger b \tilde{\rho} - \tilde{\rho} b^\dagger b),
\end{equation}
where $\tilde{\rho}$ is the density operator of the qubits and the pseudomode, $V=W(\sigma^A_{+} + \sigma^B_{+}) b + \mathrm{H.c.}$ is the interaction Hamiltonian, and $b$ ($b^\dagger$) is the annihilation (creation) operator of the pseudomode. With $\tilde{\rho}$ at hand, $\rho_{AB}$ can be obtained via partial tracing over the pseudomode. In this way we performed calculations for the initial states $|\Phi^\pm\rangle=(|11\rangle \pm |00\rangle)/\sqrt{2}$, and it was found that both the NAQI and DIA can also be preserved for a period of time. Moreover, an appreciable amount of DIA can be generated from the initial state $|11\rangle$ (e.g., $F_a^{\max}\approx 0.7358$ for $R=10$), but there is no NAQI being generated in this case. It is expected that this dilemma may be overcame by introducing frequency detunings to the two qubits.

One may also be concerned with the case that the two qubits are immersed in two independent reservoirs. Assume that the two reservoirs are the same, then $H_\mathrm{int}$ can still be given by Eq. \eqref{eq3-1}, while $H_{SR} = H_{SR,A}+H_{SR,B}$, where
\begin{equation}\label{eq6-2}
 H_{SR,n}= \omega_n \sigma^n_{+} \sigma^n_{-} + \sum_k \omega_k b_k^\dagger b_k \hspace{0.6em} (n=A,B),
\end{equation}
and for the Lorentzian spectrum in Eq. \eqref{eq3-2}, by fixing $\alpha_A=\alpha_B$ and $\delta_A=\delta_B\equiv \delta$, we performed calculation for the initial state $|\Psi^+\rangle$ using the analytical formulas obtained in Refs. \cite{indep,detun}. The results showed that both in the bad and good cavity limits, the NAQI and DIA can be preserved for a longer time than that in the common reservoir. Moreover, the degree of the preserved NAQI and DIA can be further improved by enlarging $\delta$, the effect of which is similar to what we have seen for the common reservoir. But in this case, one cannot achieve NAQI and DIA generation owing to the absence of the reservoir-induced interaction between the two qubits.

The strategy for preserving NAQI and DIA discussed above is in fact a ``passive" strategy, as the key idea is to suppress decoherence by optimizing the system and reservoir parameters. In addition, one can resort to some active strategies to achieve a protection of NAQI and DIA well beyond their natural decay times. For example, for the common reservoir considered in this work, if the detunings $\delta_A=\delta_B$, then by using the measurement-induced quantum Zeno effect presented in Ref. \cite{common1}, one can protect actively the NAQI and DIA well. In fact, for the initial state $|\Psi^+\rangle$, by performing the prescribed measurements at time intervals $T$, one can obtain its survival probability as
\begin{equation}\label{eq6-3}
 P^{(N)}(t)= |\beta_{+}^2 p(t) +\beta_{-}^2 |^{2N},
\end{equation}
where $N=t/T$ is the number of measurements in a finite time $t$ and $\beta_{\pm}=(r_B \pm r_A)/\sqrt{2}$. Denoting $\gamma_0(T)=-\ln[|\beta_{+}^2 p(T)+\beta_{-}^2 |^2]/T$, then $P^{(N)}(t)= e^{-\gamma_0(T) t}$, which approaches 1 at $N\rightarrow \infty$ for a finite $t$. This shows that when being measured frequently, the decay rate of the system will be effectively suppressed as it is reset to the initial state with a high probability after each prescribed measurement. Thereby, a Zeno protection of both the NAQI and DIA is achieved

Finally, it is also worthy to further consider the finite-temperature reservoirs and to extend the investigation to multipartite systems. For two independent reservoirs in the Markovian regime \cite{newref2}, we performed calculation by solving the master equation in the Lindblad form. The result showed that both the NAQI and DIA degrade with the increasing temperature (or equivalently, with the increasing average thermal photons initially in the reservoir), but a general study on the details of the finite-temperature effects and non-Markovianity of the common reservoir on NAQI and DIA is still needed. As for the scalability of NAQI and DIA as network resources or the shareability of them among multipartite systems, one can show that for certain multipartite states (e.g., the prototype \textit{W} state), the assisted imaginarity distillation task can have a quantum advantage among more than two qubits. But due to its hierarchy with steerability \cite{naqi}, the NAQI can be shared by at most two qubits, that is, a party $A$ cannot have NAQI with two parties $B$ and $C$ simultaneously. Of course, this does not rule out the possibility of sharing NAQI between qubits $A$ and $B$ and entanglement between qubits $A$ and $C$. For example, for the class of the \textit{W} states $|W_n\rangle_{123}= (|100\rangle+ \sqrt{n}|010\rangle+ \sqrt{n+1}|001\rangle)/ \sqrt{2n+2}$ \cite{newref3}, there is entanglement between any two qubits and NAQI between qubits 2 and 3 when $n \gtrsim 1.25$.

\section*{ACKNOWLEDGMENTS}
This work was supported by the National Natural Science Foundation of China (Grants No. 12275212, No. T2121001, No. 92265207, and No. 92365301) and the Youth Innovation Team of Shaanxi Universities (Grant No. 24JP177).

\section*{DATA AVAILABILITY}
The data that support the findings of this article are openly available \cite{data}.

\begin{appendix}

\section{Upper bound of geometric imaginarity with respect to the MUBs} \label{sec:A}
\setcounter{equation}{0}
\renewcommand{\theequation}{A\arabic{equation}}

For the one-qubit state $\rho= (\iden + \bm{b} \cdot \bm{\sigma})/2$ with $\bm{b} = (b_1,b_2,b_3)$ and $|\bm{b}|\leqslant 1$, in the reference basis $\mathcal{Z}$ spanned by the eigenbasis of $\sigma_3$, one can obtain the eigenvalues of $\rho^{1/2}\rho^T\rho^{1/2}$ as
\begin{equation}\label{eqa-1}
 \lambda_{1,2}= \frac{1}{4}\left(\sqrt{1-b_2^2} \pm \sqrt{b_1^2+b_3^2} \right)^2,
\end{equation}
where $\lambda_1$ takes the ``$+$" sign and $\lambda_2$ takes the ``$-$" sign. This, together with Eq. \eqref{eq2-2}, yields
\begin{equation}\label{eqa-2}
 \mathcal{I}_g^{\mathcal{Z}}(\rho) = \frac{1}{2} \left(1-\sqrt{1-b_2^2}\right).
\end{equation}

Similarly, in the reference bases $\mathcal{X}$ and $\mathcal{Y}$ spanned by the eigenbases of $\sigma_1$ and $\sigma_2$, respectively, one has
\begin{equation}\label{eqa-3}
\begin{aligned}
 \mathcal{I}_g^{\mathcal{X}}(\rho) = \mathcal{I}_g^{\mathcal{Z}}(\rho),~
 \mathcal{I}_g^{\mathcal{Y}}(\rho) = \frac{1}{2} \left(1-\sqrt{1-b_1^2}\right),
\end{aligned}
\end{equation}
and it follows that
\begin{equation}\label{eqa-4}
 \sum_{i=\mathcal{X},\mathcal{Y},\mathcal{Z}} \mathcal{I}_g^{i}(\rho) = \frac{1}{2} \left(3-2\sqrt{1-b_2^2}-\sqrt{1-b_1^2}\right) \leqslant 1,
\end{equation}
where the equality holds when $b_1 = 0$ and $b_2 = 1$. As maximizing $\sum_{i=\mathcal{X},\mathcal{Y},\mathcal{Z}} \mathcal{I}_g^{i}(\rho)$ over all the unitary equivalent classes of $\{\mathcal{X},\mathcal{Y},\mathcal{Z}\}$ is equivalent to maximizing it over all the unitary equivalent states of $\rho$, one has $\mathcal{I}_{g,\mathcal{B}}=1$.

\section{Calculation of $\mathcal{N}_\alpha(\rho_{AB})$} \label{sec:B}
\setcounter{equation}{0}
\renewcommand{\theequation}{B\arabic{equation}}

For $\rho_{AB}$ obtained from Eq. \eqref{eq3-4}, after Alice performing the local projective measurement $\Pi_i^\pm=(\iden \pm \bm{n}_i\cdot \bm{\sigma})/2$ on qubit $A$, the conditional state of Bob in Eq. \eqref{eq2-4} can be written as
\begin{equation} \label{eqb-1}
 \rho_{B|\Pi_i^a}= \frac{1}{2}(\iden + \bm{b}^a_i \cdot \bm{\sigma}),
\end{equation}
where $\bm{b}^a_i= (b^a_{i,1}, b^a_{i,2}, b^a_{i,3})$ is the local Bloch vector with the elements being given by
\begin{equation} \label{eqb-2}
\begin{aligned}
 & b^\pm_{i,1}= \frac{\pm \mathrm{Re}\big[\rho_{AB}^{2,3} e^{i\phi_i}\big] \sin\theta_i}{p_{B|\Pi_i^\pm}}, ~
   b^\pm_{i,2}= \frac{\pm \mathrm{Im}\big[\rho_{AB}^{2,3} e^{i\phi_i}\big] \sin\theta_i}{p_{B|\Pi_i^\pm}}, \\
 & b^\pm_{i,3}= \frac{\rho_{AB}^{3,3}(1 \mp \cos\theta_i)}{p_{B|\Pi_i^\pm}} -1,
\end{aligned}
\end{equation}
and $\rho_{AB}^{i,j}$ represents the element in the $i$th row and $j$th column of $\rho_{AB}$, while the probability of the measurement outcome $a\in\{+,-\}$ is given by
\begin{equation} \label{eqb-3}
 p_{B|\Pi_i^\pm}=\frac{1}{2}(1\mp \cos\theta_i) \pm\rho_{AB}^{2,2}\cos\theta_i.
\end{equation}

With $\rho_{B|\Pi_i^a}$ of Eq. \eqref{eqb-1} at hand, we further rewrite it in the reference basis $\{|\mathcal{M}_i^{\pm}\rangle\}$ [see Eq. \eqref{eq2-6}]. This, together with Eqs. \eqref{eq2-2} and \eqref{eqb-3}, yields $\sum_{i,a} p_{B|\Pi_i^a} \mathcal{I}_\alpha^{\mathcal{M}_i}(\rho_{B|\Pi_i^a})$, and then optimizing it numerically over $\{\Pi_i,\mathcal{M}_i\}$, one can obtain $\mathcal{N}_\alpha(\rho_{AB})$.

\end{appendix}

\newcommand{\AdP}{Ann. Phys. (Berlin) }
\newcommand{\APL}{Appl. Phys. Lett. }
\newcommand{\AoP}{Ann. Phys. (N.Y.) }
\newcommand{\AQT}{Adv. Quantum Technol. }
\newcommand{\CMP}{Commun. Math. Phys. }
\newcommand{\CP}{Commun. Phys. }
\newcommand{\CTP}{Commun. Theor. Phys. }
\newcommand{\EPL}{Europhys. Lett. }
\newcommand{\EPJD}{Eur. Phys. J. D }
\newcommand{\EPJP}{Eur. Phys. J. Plus }
\newcommand{\IJTP}{Int. J. Theor. Phys. }
\newcommand{\JPA}{J. Phys. A }
\newcommand{\JPB}{J. Phys. B }
\newcommand{\JMP}{J. Math. Phys. }
\newcommand{\NC}{Nat. Commun. }
\newcommand{\NJP}{New J. Phys. }
\newcommand{\NP}{Nat. Phys. }
\newcommand{\PR}{Phys. Rep. }
\newcommand{\PRL}{Phys. Rev. Lett. }
\newcommand{\PRA}{Phys. Rev. A }
\newcommand{\PRB}{Phys. Rev. B }
\newcommand{\PRD}{Phys. Rev. D }
\newcommand{\PRE}{Phys. Rev. E }
\newcommand{\PRX}{Phys. Rev. X }
\newcommand{\PRR}{Phys. Rev. Res. }
\newcommand{\PRAp} {Phys. Rev. Appl. }
\newcommand{\PLA}{Phys. Lett. A }
\newcommand{\PA}{Physica A }
\newcommand{\PS}{Phys. Scr. }
\newcommand{\QIP}{Quantum Inf. Process. }
\newcommand{\RMP}{Rev. Mod. Phys. }
\newcommand{\RPP}{Rep. Prog. Phys. }
\newcommand{\SCPMA}{Sci. China-Phys. Mech. Astron. }
\newcommand{\SR}{Sci. Rep. }


\begin{thebibliography}{50}


\bibitem{real1} M.-O. Renou, D. Trillo, M. Weilenmann, T. P. Le, A. Tavakoli, N. Gisin, A. Ac\'{\i}n, and M. Navascu{\'e}s, Quantum theory based on real numbers can be experimentally falsified, \href{https://doi.org/10.1038/s41586-021-04160-4} {Nature (London) \textbf {600}, 625 (2021)}.

\bibitem{real2} N. Prasannan, S. De, S. Barkhofen, B. Brecht, C. Silberhorn, and J. Sperling, Experimental entanglement characterization of two-rebit states, \href{https://doi.org/10.1103/PhysRevA.103.L040402} {\PRA \textbf{103}, L040402 (2021)}.

\bibitem{real3} Z.-D. Li, Y.-L. Mao, M. Weilenmann, A. Tavakoli, H. Chen, L. Feng, S.-J. Yang, M.-O. Renou, D. Trillo, T. P. Le, N. Gisin, A. Acin, M. Navascues, Z.Wang, and J. Fan, Testing real quantum theory in an optical quantum network, \href{https://doi.org/10.1103/PhysRevLett.128.040402} {\PRL \textbf{128}, 040402 (2022)}.

\bibitem{real4} M.-C. Chen, C. Wang, F.-M. Liu, J.-W. Wang, C. Ying, Z.-X. Shang, Y. Wu, M. Gong, H. Deng, F. T. Liang, Q. Zhang, C. Z. Peng, X. Zhu, A. Cabello, C. Y. Lu, and J. W. Pan, Ruling out real-valued standard formalism of quantum theory, \href{https://doi.org/10.1103/PhysRevLett.128.040403} {\PRL \textbf{128}, 040403 (2022)}.

\bibitem{real5} A. Bednorz and J. Batle, Optimal discrimination between real and complex quantum theories, \href{https://doi.org/10.1103/PhysRevA.106.042207} {\PRA \textbf{106}, 042207 (2022)}.

\bibitem{QE0} R. Horodecki, P. Horodecki, M. Horodecki, and K. Horodecki, Quantum entanglement, \href{https://doi.org/10.1103/RevModPhys.81.865}{\RMP \textbf{81}, 865 (2009)}.

\bibitem{QE1} V. Vedral, M. B. Plenio, M. A. Rippin, and P. L. Knight, Quantifying entanglement, \href{https://doi.org/10.1103/PhysRevLett.78.2275}{\PRL \textbf{78}, 2275 (1997)}.

\bibitem{coher} T. Baumgratz, M. Cramer, and M. B. Plenio, Quantifying coherence, \href{https://doi.org/10.1103/PhysRevLett.113.140401} {\PRL \textbf {113}, 140401 (2014)}.

\bibitem{Plenio} A. Streltsov, G. Adesso, and M. B. Plenio, Colloquium: Quantum coherence as a resource, \href{https://doi.org/10.1103/RevModPhys.89.041003} {\RMP \textbf{89}, 041003 (2017)}.

\bibitem{Hu} M. L. Hu, X. Hu, J. C. Wang, Y. Peng, Y. R. Zhang, and H. Fan, Quantum coherence and geometric quantum discord, \href{https://doi.org/10.1016/j.physrep.2018.07.004} {\PR \textbf{762--764}, 1 (2018)}.


\bibitem{Rsuper} T. Theurer, N. Killoran, D. Egloff, and M. B. Plenio, Resource theory of superposition, \href{https://doi.org/10.1103/PhysRevLett.119.230401} {\PRL\textbf {119}, 230401 (2017)}.

\bibitem{Rsteer} R. Gallego and L. Aolita, Resource theory of steering, \href{https://doi.org/10.1103/PhysRevX.5.041008} {\PRX \textbf{5}, 041008 (2015)}.

\bibitem{ima_jpa} A. Hickey and G. Gour, Quantifying the imaginarity of quantum mechanics, \href{https://doi.org/10.1088/1751-8121/aabe9c} {\JPA \textbf{51}, 414009 (2018)}.

\bibitem{ima_prl} K. D. Wu, T. V. Kondra, S. Rana, C. M. Scandolo, G. Y. Xiang, C. F. Li, G. C. Guo, and A. Streltsov, Operational resource theory of imaginarity, \href{https://doi.org/10.1103/PhysRevLett.126.090401} {\PRL \textbf{126}, 090401 (2021)}.

\bibitem{ima_pra} K. D. Wu, T. V. Kondra, S. Rana, C. M. Scandolo, G. Y. Xiang, C. F. Li, G. C. Guo, and A. Streltsov, Resource theory of imaginarity: Quantification and state conversion, \href{https://doi.org/10.1103/PhysRevA.103.032401} {\PRA \textbf{103}, 032401 (2021)}.

\bibitem{ima_njp}T. V. Kondra, C. Datta, and A. Streltsov, Real quantum operations and state transformations, \href{https://doi.org/10.1088/1367-2630/acf9c4} {\NJP \textbf{25}, 093043 (2023)}.

\bibitem{ima_qip} S. Xue, J. Guo, P. Li, M. Ye, and Y. Li, Quantification of resource theory of imaginarity, \href{https://doi.org/10.1007/s11128-021-03324-5} {\QIP \textbf{20}, 383 (2021)}.

\bibitem{ima_dus} S. Du and Z. Bai, Quantifying imaginarity in terms of pure-state imaginarity, \href{https://doi.org/10.1103/PhysRevA.111.022405} {\PRA \textbf{111}, 022405 (2025)}.

\bibitem{ima_lin} L. Zhang and N. Li, Quantifying imaginarity via the quantum optimal transport cost, \href{https://doi.org/10.1140/epjp/s13360-025-06592-7} {\EPJP \textbf{140}, 669 (2025)}.

\bibitem{ima_epsilon} Y. Luo, F. Meng, and Y. Wang, Epsilon measures of state-based quantum resource theory, \href{https://doi.org/10.1103/PhysRevA.109.052413} {\PRA \textbf{109}, 052413 (2024)}.


\bibitem{ima_suny} P. Tian and Y. Sun, Generalized quantum Jensen-Shannon divergence of imaginarity, \href{https://doi.org/10.1016/j.physleta.2025.130479} {\PLA \textbf{544}, 130479 (2025)}.

\bibitem{ima_pla} J. Xu, Quantifying the imaginarity of quantum states via Tsallis relative entropy, \href{https://doi.org/10.1016/j.physleta.2024.130024} {\PLA \textbf{528}, 130024 (2024)}.

\bibitem{ima_aqt} M. L. Guo, S. Y. Huang, B. Li, and S. M. Fei, Quantifying the imaginarity via different distance measures, \href{https://doi.org/10.1002/qute.202400562} {\AQT \textbf{8}, 2400562 (2025)}.

\bibitem{ima_maxmin} L. Zhang and N. Li, Maximum and minimum relative entropies of imaginarity, \href{https://doi.org/10.1007/s11128-025-04983-4} {\QIP \textbf{24}, 361 (2025)}.

\bibitem{ima_ctp} C. Wu and Z. Wu, Two imaginarity monotones induced by unified ($\alpha,\beta$)-relative entropy, \href{https://doi.org/10.1088/1572-9494/adbdbd} {\CTP \textbf{77}, 095101 (2025)}.

\bibitem{ima_qc1} H. B. Li, M. Hua, Q. Zheng, Q. J. Zhi, and Y. Ping, Relationship between robustness of imaginarity and quantum coherence, \href{https://doi.org/10.1140/epjd/s10053-023-00618-4} {\EPJD \textbf{77}, 38 (2023)}.

\bibitem{ima_qc2} L. Zhang and N. Li, Coherence as maximal imaginarity generated by incoherent operations, \href{https://doi.org/10.1209/0295-5075/ad847d} {\EPL \textbf{148}, 28002 (2024)}.

\bibitem{ima_qc3} J. Xu, Coherence and imaginarity of quantum states, \href{https://doi.org/10.1088/1402-4896/ad99a1} {\PS \textbf{100}, 025106 (2025)}.

\bibitem{ima_ent1} Y. Sun, R. Ren, Y. Wang, and Y. Li, Analysis of the relationship between imaginarity and entanglement, \href{https://doi.org/10.1103/PhysRevA.111.032425} {\PRA \textbf{111}, 032425 (2025)}.

\bibitem{ima_ent2} J. Zhang, Y. Luo, and Y. Li, Imaginaring and deimaginaring power of quantum channels and the trade-off between imaginarity and entanglement, \href{https://doi.org/10.1007/s11128-023-04131-w} {\QIP \textbf{22}, 405 (2023)}.


\bibitem{ima_speed} D. P. Xuan, Z. X. Shen, W. Zhou, S. M. Fei, and Z. X. Wang, Quantum-imaginarity-based quantum speed limit, \href{https://doi.org/10.1103/s7kr-8rrn} {\PRA \textbf{112}, 052202 (2025)}.

\bibitem{ima_erase} X. Shi, Erasing imaginarity: An operational method, \href{https://doi.org/10.1103/PhysRevA.111.L050401} {\PRA \textbf{111}, L050401 (2025)}.

\bibitem{ima_wit1} C. Fernandes, R. Wagner, L. Novo, and E. F. Galvao, Unitary-invariant witnesses of quantum imaginarity, \href{https://doi.org/10.1103/PhysRevLett.133.190201} {\PRL \textbf{133}, 190201 (2024)}.

\bibitem{ima_wit2} L. Zhang and N. Li, On imaginarity witnesses, \href{https://doi.org/10.1016/j.physleta.2024.130135} {\PLA \textbf{530}, 130135 (2025)}.

\bibitem{ima_gauss1} J. Xu, Imaginarity of Gaussian states, \href{https://doi.org/10.1103/PhysRevA.108.062203} {\PRA \textbf{108}, 062203 (2023)}.

\bibitem{ima_gauss2} T. Zhang, J. Hou, and X. Qi, An easily computable measure of Gaussian quantum imaginarity, \href{https://doi.org/10.1002/andp.202500171}{\AdP \textbf{537}, e2500171 (2025)}.

\bibitem{ima_channel} X. Chen and Q. Lei, Imaginarity of quantum channels: Refinement and alternative, \href{https://doi.org/10.1016/j.physleta.2024.130129} {\PLA \textbf{530}, 130129 (2025)}.

\bibitem{ima_mask1} R. Q. Zhang, Z. Hou, Z. Li, H. Zhu, G. Y. Xiang, C. F. Li, and G. C. Guo, Experimental masking of real quantum states, \href{https://doi.org/10.1103/PhysRevApplied.16.024052} {\PRAp \textbf{16}, 024052 (2021)}.

\bibitem{ima_mask2} H. Zhu, Hiding and masking quantum information in complex and real quantum mechanics, \href{https://doi.org/10.1103/PhysRevResearch.3.033176}{\PRR \textbf{3}, 033176 (2021)}.

\bibitem{ima_estim} J. Miyazaki and K. Matsumoto, Imaginarity-free quantum multiparameter estimation, \href{https://doi.org/10.22331/q-2022-03-10-665} {Quantum \textbf{6}, 665 (2022)}.


\bibitem{ima_broad1} Z. Zhang, N. Li, and S. Luo, Broadcasting of imaginarity, \href{https://doi.org/10.1103/PhysRevA.110.052439} {\PRA \textbf{110}, 052439 (2024)}.

\bibitem{ima_broad2} L. Zhang and N. Li, Can imaginarity be broadcast via real operations?, \href{https://doi.org/10.1088/1572-9494/ad6de5} {\CTP \textbf{76}, 115104 (2024)}.

\bibitem{naqi} Z. W. Wei and S. M. Fei, Nonlocal advantages of quantum imaginarity, \href{https://doi.org/10.1103/PhysRevA.110.052202} {\PRA \textbf{110}, 052202 (2024)}.

\bibitem{naqc1} D. Mondal, T. Pramanik, and A. K. Pati, Nonlocal advantage of quantum coherence, \href{https://doi.org/10.1103/PhysRevA.95.010301} {\PRA \textbf{95}, 010301 (2017)}.

\bibitem{naqc2} M. L. Hu and H. Fan, Nonlocal advantage of quantum coherence in high-dimensional states, \href{https://doi.org/10.1103/PhysRevA.98.022312} {\PRA \textbf{98}, 022312 (2018)}.

\bibitem{naqc3} M. L. Hu, X. M. Wang, and H. Fan, Hierarchy of the nonlocal advantage of quantum coherence and Bell nonlocality, \href{https://doi.org/10.1103/PhysRevA.98.032317} {\PRA \textbf{98}, 032317 (2018)}.

\bibitem{ima_cp} K. D. Wu, T. V. Kondra, C. M. Scandolo, S. Rana, G. Y. Xiang, C. F. Li, G. C. Guo, and A. Streltsov, Resource theory of imaginarity in distributed scenarios, \href{https://doi.org/10.1038/s42005-024-01649-y} {\CP \textbf{7}, 171 (2024)}.

\bibitem{dist_coher} E. Chitambar, A. Streltsov, S. Rana, M. N. Bera, G. Adesso, and M. Lewenstein, Assisted distillation of quantum coherence, \href{https://doi.org/10.1103/PhysRevLett.116.070402} {\PRL \textbf{116}, 070402 (2016)}.

\bibitem{ima_dy1} T. Xia, J. Xu, and M. J. Zhao, A note on geometric imaginarity, \href{https://doi.org/10.1007/s10773-024-05866-7} {\IJTP \textbf{63}, 319 (2024)}.

\bibitem{ima_free} B. Zheng, Z. Guo, C. Zhang, and H. Cao, Freezing imaginarity measures with real operations, \href{https://doi.org/10.1088/1751-8121/ada8e8} {\JPA \textbf{58}, 035304 (2025)}.


\bibitem{common1} S. Maniscalco, F. Francica, R. L. Zaffino, N. L. Gullo, and F. Plastina, Protecting entanglement via the quantum Zeno effect, \href{https://doi.org/10.1103/PhysRevLett.100.090503} {\PRL \textbf{100}, 090503 (2008)}.

\bibitem{common2} F. Francica, S. Maniscalco, J. Piilo, F. Plastina, and K.-A. Suominen, Off-resonant entanglement generation in a lossy cavity, \href{https://doi.org/10.1103/PhysRevA.79.032310} {\PRA \textbf{79}, 032310 (2009)}.

\bibitem{common0} Y. X. Xie, J. Wang, and Y. Liu, Performance enhancement of a quantum battery via frequency detuning, \href{https://doi.org/10.1088/1402-4896/adf51d} {\PS \textbf{100}, 085112 (2025)}.

\bibitem{nature1} A. Wallraff, D. I. Schuster, A. Blais, L. Frunzio, R. S. Huang, J. Majer, S. Kumar, S. M. Girvin, and R. J. Schoelkopf, Strong coupling of a single photon to a superconducting qubit using circuit quantum electrodynamics, \href{https://doi.org/10.1038/nature02851}{Nature (London) \textbf{431}, 162 (2004)}.

\bibitem{nature2} M. A. Sillanp\"{a}\"{a}, J. I. Park, and R. W. Simmonds, Coherent quantum state storage and transfer between two phase qubits via a resonant cavity, \href{https://doi.org/10.1038/nature06124} {Nature (London) \textbf{449}, 438 (2007)}.

\bibitem{nature3} J. Majer, J. M. Chow, J. M. Gambetta, J. Koch, B. R. Johnson, J. A. Schreier, L. Frunzio, D. I. Schuster, A. A. Houck, A. Wallraff, A. Blais, M. H. Devoret, S. M. Girvin, and R. J. Schoelkopf, Coupling superconducting qubits via a cavity bus, \href{https://doi.org/10.1038/nature06184} {Nature (London) \textbf{449}, 443 (2007)}.

\bibitem{apl} S. Kuhr, S. Gleyzes, C. Guerlin, J. Bernu, U. B. Hoff, S. Del\'{e}glise, S. Osnaghi, M. Brune, J.-M. Raimond, S. Haroche, E. Jacques, P. Bosland, and B. Visentin, Ultrahigh finesse Fabry-P\'{e}rot superconducting resonator, \href{http://doi.org/10.1063/1.2724816} {\APL \textbf{90}, 164101 (2007)}.

\bibitem{newref1} Y. X. Xiao, D. Feng, X. Y. Gu,  G. H. Liang, M. C. Wang, Z. Y. Peng, B. J. Chen, Y. Yan, Z. Y. Mei, S. L. Zhao, Y. Z. Bu, C. L. Deng, K. Yang, Y. Tian, X. Song, D. Zheng, Y. X. Zhang, Y. H. Shi, Z. Xiang, K. Xu, and H. Fan, Flexible readout and unconditional reset for superconducting multiqubit processors with tunable Purcell filters, \href{https://doi.org/10.1103/vwrv-x1kr} {\PRL \textbf{136}, 070601 (2026)}.

\bibitem{pmode1} B. M. Garraway, Nonperturbative decay of an atomic system in a cavity, \href{https://doi.org/10.1103/PhysRevA.55.2290} {\PRA \textbf{55}, 2290 (1997)}.

\bibitem{pmode2} B. J. Dalton, S. M. Barnett, and B. M. Garraway, Theory of pseudomodes in quantum optical processes, \href{https://doi.org/10.1103/PhysRevA.64.053813} {\PRA \textbf{64}, 053813 (2001)}.


\bibitem{pmode3} B. J. Dalton and B. M. Garraway, Non-Markovian decay of a three-level cascade atom in a structured reservoir, \href{https://doi.org/10.1103/PhysRevA.68.033809} {\PRA \textbf{68}, 033809 (2003)}.

\bibitem{pmode4} L. Mazzola, S. Maniscalco, J. Piilo, K.-A. Suominen, and B. M. Garraway, Sudden death and sudden birth of entanglement in common structured reservoirs, \href{https://doi.org/10.1103/PhysRevA.79.042302} {\PRA \textbf{79}, 042302 (2009)}.

\bibitem{pmode5} D. Z. Rossato, T. Werlang, L. K. Castelano, C. J. Villas-Boas, and F. F. Fanchini, Purity as a witness for initial system-environment correlations in open-system dynamics, \href{https://doi.org/10.1103/PhysRevA.84.042113} {\PRA \textbf{84}, 042113 (2011)}.

\bibitem{common3} L. Mazzola, S. Maniscalco, J. Piilo, and K.-A. Suominen, Interplay between entanglement and entropy in two-qubit systems, \href{https://doi.org/10.1088/0953-4075/43/8/085505} {\JPB \textbf{43}, 085505 (2010)}.

\bibitem{indep} B. Bellomo, R. L. Franco, and G. Compagno, Non-Markovian effects on the dynamics of entanglement, \href{http://doi.org/10.1103/PhysRevLett.99.160502} {\PRL \textbf{99}, 160502 (2007)}.

\bibitem{detun} M. L. Hu and H. Fan, Quantum-memory-assisted entropic uncertainty principle, teleportation, and entanglement witness in structured reservoirs, \href{https://doi.org/10.1103/PhysRevA.86.032338} {\PRA \textbf{86}, 032338 (2012)}.

\bibitem{newref2} S. Fedortchenko, A. Keller, T. Coudreau, and P. Milman, Finite-temperature reservoir engineering and entanglement dynamics, \href{https://doi.org/10.1103/PhysRevA.90.042103} {\PRA \textbf{90}, 042103 (2014)}.

\bibitem{newref3} P. Agrawal and A. Pati, Perfect teleportation and superdense coding with \textit{W} states, \href{https://doi.org/10.1103/PhysRevA.74.062320}{\PRA \textbf{74}, 062320 (2006)}.


\bibitem{data} S. M. Wang, M. L. Hu, and H. Fan, Data for ``Off-resonant preservation and generation of imaginarity in distributed scenarios" [Data set], Zenodo (2026), doi: \href{https://doi.org/10.5281/zenodo.19412242} {10.5281/zenodo.19412242}.


\end{thebibliography}

\end{document}